\documentclass[11pt]{article}
\usepackage{url}

\usepackage{natbib, graphicx,setspace,lscape,longtable}
\usepackage{epsfig}
\usepackage{amssymb, amsmath,bm,color}
\usepackage{multirow}

\usepackage{xcolor}         
\usepackage{amsmath}                        
\usepackage{graphicx}               
\usepackage{caption}
\usepackage{subcaption, fullpage}
\usepackage{hyperref}
\usepackage[english]{babel}

\definecolor{DarkGreen}{rgb}{0.5,0.8,0.6}   
\definecolor{RGBblack}{rgb}{0.0,0.0,0.0}    

\definecolor{grau}{rgb}{0.8,0.8,0.8}

\newcommand{\chen}[1]{\color{orange}}

\newcommand{\is}{\itemsep=0pt}
\newcommand{\bd}[1]{\begin{description}[#1]\is}
  \newcommand{\ed}{\end{description}}
\newcommand{\bi}{\begin{itemize}\is}
  \newcommand{\ei}{\end{itemize}}
\newcommand{\be}{\begin{enumerate}\is}
  \newcommand{\ee}{\end{enumerate}}
  \newcommand{\beq}{\begin{eqnarray}\is}
  \newcommand{\eeq}{\end{eqnarray}}

\makeatletter
\newcommand*{\rom}[1]{\expandafter\@slowromancap\romannumeral #1@}

\newcommand{\bSigma}{\bm{\Sigma}}
\newcommand{\bLambda}{\bm{\Lambda}}
\newcommand{\bPsi}{\bm{\Psi}}
\newcommand{\btheta}{\bm{\theta}}

\newcommand{\bv}{\bm{v}}

\newcommand{\bmu}{\bm{\mu}}
\newcommand{\bbeta}{\bm{\beta}}

\newcommand{\bY}{\bm{Y}}
\newcommand{\bO}{\bm{O}}
\newcommand{\bX}{\bm{X}}
\newcommand{\bV}{\bm{V}}

\newcommand{\bR}{\bm{R}}
\newcommand{\bU}{\bm{U}}
\newcommand{\bI}{\bm{I}}
\newcommand{\bS}{\bm{S}}
\renewcommand{\th}{\theta}

\newcommand{\bths}{\bm{\th^\star}}

\newcommand{\bx}{\bm{x}}

\newcommand{\bth}{\bm \theta}

\newcommand{\Sig}{\Sigma}
\newcommand{\bSig}{\bSigma}

\newcommand{\Bern}{\mbox{Bern}}
\newcommand{\Unif}{\mbox{Unif}}

\newcommand{\GP}{\mbox{GP}}
\newcommand{\DP}{\mbox{DP}}

\newcommand{\DDPGP}{\mbox{DDP-GP}}
\newcommand{\IW}{\mbox{Inverse-Wishart}}

\newcommand{\Yp}{Y_P}
\newcommand{\Ypi}{Y_{Pi}}
\newcommand{\Yd}{Y_D}
\newcommand{\Ydi}{Y_{Di}}

\begin{document}
\doublespacing

\title{A Bayesian Nonparametric Approach for Evaluating the Causal Effect of Treatment in Randomized Trials with Semi-Competing Risks}

\author{Yanxun Xu$^{1}$, Daniel Scharfstein$^{2}$, Peter M\"uller$^{3}$, Michael  Daniels$^{4}$ \\
$^{1}$Department of Applied Mathematics and Statistics, \\
Johns Hopkins University. yanxun.xu@jhu.edu\\
$^{2}$Department of Biostatistics, Johns Hopkins University. dscharf@jhu.edu\\
$^{3}$Department of Mathematics, The University of Texas at Austin. pmueller@math.utexas.edu\\ 
$^{4}$Department of Statistics, University of Florida. daniels@ufl.edu\\ 
}

\date{}
\maketitle


\begin{abstract}
We develop a Bayesian nonparametric (BNP) approach to evaluate the causal effect of treatment in a randomized trial where a nonterminal event may be censored by a terminal event, but not vice versa (i.e., semi-competing risks).   Based on the idea of principal stratification, we define a novel estimand for the causal effect of treatment on the nonterminal event. We introduce identification assumptions, indexed by a sensitivity parameter, and show how to draw inference using our BNP approach.  We conduct simulation studies and illustrate our methodology using data from a brain cancer trial.

\noindent{\bf KEY WORDS:}   
{Bayesian nonparametrics; Brain cancer trial; Causal inference; Identification assumptions; Principal stratification; Sensitivity analysis.}
\end{abstract}


\section{Introduction}
\label{sec:intro}
Semi-competing risks \citep{fine2001semi} occur in studies where observation of a nonterminal event (e.g., progression) may be pre-empted by a terminal event (e.g., death), but not vice versa. 
In randomized clinical trials to evaluate treatments of life-threatening diseases, patients are often observed for specific types of disease progression and survival. Often, 
the primary outcome is patient survival, resulting in data analyses focusing on the terminal event using standard survival analysis tools \citep{ibrahim2005bayesian}.  However, there may also be interest in understanding the causal effect of treatment on nonterminal outcomes such as progression, readmission, etc. 
An example is a randomized trial for the treatment of malignant brain tumors, where one of the important progression endpoints is based on deterioration of the cerebellum. An important feature of this progression endpoint is that it is biologically plausible that a patient could die without cerebellar deterioration. Thus, analyzing the effect of treatment on progression needs to account for the fact that progression is not well-defined after death.

\cite{varadhan2014semicompeting} reviews models that have been proposed for analyzing semi-competing data.  These models can be classified into two broad categories: models for the distribution of the observable data, e.g., cause-specific hazards, sub-distribution functions  \citep{fix1951simple, hougaard1999multi, xu2010statistical,lee2015bayesian} and models for the distribution of the latent failure times  \citep{robins1995analytic1, robins1995analytic2, lin1996comparing, wang2003estimating, peng2007regression, adam2009marginal, peng2012rank, chen2012maximum,  hsieh2012regression, comment2019survivor}.    \cite{xu2010statistical}  argued against the use of latent failure time models because the marginal distribution of the nonterminal event is hypothetical.  This is because the joint distribution of the nonterminal event ($Y_P$) and terminal event ($Y_D$) is only identified on a wedge of $R^2$. Rather, they argued that ``semi-competing risks data are better modeled using an illness-death compartment model", where ``a subject can either transit directly to the terminal event or first to the nonterminal event and then to the terminal event."  They proposed a Markov shared frailty model for the transition rates.  \cite{lee2015bayesian} proposed a Bayesian semiparametric extension, which focused on estimation of regression parameters, characterization of dependence between event times and prediction of event times for specific covariate profiles.
The latent failure approaches of \cite{fine2001semi}, \cite{wang2003estimating} and \cite{peng2007regression} have focused on estimating regression parameters and estimating dependence between nonterminal and terminal event times using copula models.  \cite{robins1995analytic1, robins1995analytic2}  focused solely on estimating regression parameters and discusses causal interpretability.  Recently, \cite{comment2019survivor} proposed a casual estimand similar to the one we discuss here, but uses different models (i.e., parametric frailty models) and different causal assumptions (i.e., latent ignorability).

In this paper, we are interested in estimating the causal effect of treatment on the nonterminal endpoint from a randomized trial generating semi-competing risk data.  Using the potential outcomes framework \citep{rubin1974estimating}, we propose a principal stratification estimand \citep{frangakis2002principal} to quantify the causal effect.  Our estimand is a time-varying version of the survival average causal effect (see, e.g., \cite{zhang2003estimation, tchetgen2014identification}), quantified on a relative risk scale. 
We  introduce assumptions that utilize baseline covariates to identify this estimand from the distribution of the observable data and propose a Bayesian nonparametric (BNP) approach for modeling this distribution.  An important feature of BNP models is
their large support, allowing us to approximate essentially arbitrary
distributions \citep{ishwaran2001gibbs}. To handle covariates, our approach is based
on the dependent Dirichlet process (DDP) prior introduced by \cite{maceachern1999dependent}.

The paper is outlined as follows:  Section 2 introduces the motivating brain tumor study. 
The formal definition of the causal estimand is
introduced in Section \ref{sec:causal}.  We introduce the BNP model in Section
\ref{sec:model}. A simulation study is summarized
in Section \ref{sec:simu}. We analyze the brain tumor data in Section
\ref{sec:brain}, and conclude with brief discussion in Section
\ref{sec:con}.

\section{Motivating Brain Tumor Study}
The methodology is motivated by  a  randomized and placebo-controlled phase II trial
for 222 recurrent gliomas patients, who were scheduled for tumor
resection with recurrent malignant brain tumors
\citep{brem1995placebo}. Eligible patients had a single focus of
tumor in the cerebrum, had a Karnofsky score greater than 60, had completed
radiation therapy, had not taken nitrosoureas within 6 weeks of
enrollment, and had not had systematic chemotherapy within 4 weeks of
enrollment. 
The data include 
11 baseline prognostic measures 
and a baseline evaluation of cerebellar function. 
The former includes
age, race, Karnofsky performance score, local
vs. whole brain radiation, percent of tumor resection, previous use of
nitrosoureas, and tumor histology (glioblastoma, anapestic
astrocytoma, oligodendrolioma, or other) at implantation.  
Patients were randomized to receive surgically implanted
biodegradable polymer discs with or without 3.85\% of carmustine. The
follow-up duration was 1 year. Of the 219 patients with complete
baseline measures, 204 were observed to die and 100 were observed to
progress prior to death. Of the 15 patients who did not die, 4 were
observed to have cerebellar progression.  Our goal is to
estimate the causal effect of treatment on time to cerebellar progression.

\section{Causal Estimand and Identification Assumptions}
\label{sec:causal}

\subsection{Potential Outcomes and Causal Estimand}

Let $Y_{P}^z$, $\Yd^z$ and $C^z$ denote progression time, death time and censoring time, under
treatment $z$. Here $z=0, 1$ represents control and
treatment group, respectively.  
 All event times are log-transformed.   Fundamental to our setting is
that $Y_P^z \not > \Yd^z$ (i.e., progression cannot happen after
death).  

The causal estimand of interest is the function
\begin{equation}
         \tau(u) =
  \frac{Pr[\Yp^1< u \mid \Yd^0\geq u, \Yd^1\geq u]}
       {Pr[\Yp^0< u \mid \Yd^0\geq u, \Yd^1\geq u]},
\label{eq:infer}
\end{equation}
where 
$\tau(\cdot)$ is a smooth function of $u$. 
Among patients 
who survive to time $u$ under both treatments, this estimand contrasts the risk of progression prior to time $u$ for treatment 1 relative to treatment 0, which is a causal effect in a subgroup defined by potential outcomes. This estimand is an example of a principal stratum causal effect \citep{frangakis2002principal}. 

\subsection{Observed Data}

Let $Z$ denote treatment assignment and $\bX$ denote a vector of the baseline covariates.  Let $Y_P = Y_{P}^Z$, $\Yd = \Yd^Z$ and $C = C^Z$.  Let $T_1=\Yp \wedge \Yd \wedge C$, 
    $\delta = I(\Yp < \Yd \wedge C)$, 
    $T_2=\Yd \wedge C$, and 
    $\xi=I(\Yd <  C)$
denote the observed event times and event indicators. 
The observed data for each patient are $\bO=(T_1, T_2, \delta, \xi,
Z,\bX)$.  We assume that we observe $n$ i.i.d. copies of $\bO$.  Throughout, variables subscripted by $i$ will denote data specific to patient $i$.

\subsection{Identification Assumptions}

We introduce the following four assumptions that are sufficient for identifying our causal estimand.



\noindent {\bf Assumption 1:} Treatment is randomized, i.e.,
\[
Z \perp (Y_P^z,Y^z_D,C^z,\bX);  \; \; z=0,1,
\]
and $0 < Pr[ Z=1] < 1$.

\noindent This obviously holds by design in randomized trials as considered here.

\noindent {\bf Assumption 2:} Censoring is non-informative in the sense that
\[
C^z \perp (Y^z_P,Y^z_D)  \; \vline \; \bX=\bx; \; \; z=0,1,
\]
and $Pr[ C^z  > Y^z_P, C^z > Y^z_D | \bX = \bx] > 0$ for all $\bx$.

Let $\lambda^z_{\bx}$ and $G^z_{\bx}$ denote the conditional hazard function and conditional distribution function of $\Yd^z$ given $\bX=\bx$, respectively.  Under Assumptions 1 and 2, $\lambda^z_{\bx}$ and $G^z_{\bx}$ are identified via the following formulae: 
\[
\lambda^{z}_{\bx}(t)  = \lim_{dt \rightarrow 0} \left\{ \frac{Pr[ t  \leq T_2 <  t+dt, \xi=1 \; \vline \; T_2 \geq t, \bX=\bx, Z=z]}{dt} \right\}
\]
and
\begin{eqnarray}
G^z_{\bx}(t) = 1 - \exp \left\{  - \int_0^t \lambda^{z}_{\bx}(s)  ds \right\}.
\label{eq:G}
\end{eqnarray}
Furthermore, the conditional sub-distribution function of $Y_P^z$ given $Y_D^z$ and $\bX=\bx$, $V^z_{\bx}$, is identified via the following formula:
\begin{eqnarray}
V^z_{\bx}(s|t) 
&= &Pr[  T_1 \leq s, \delta=1 \; \vline \; T_2 = t, \xi=1, \bX=\bx, Z=z],
\label{eq:V}
\end{eqnarray}
where $s \leq t$.
Together $G^z_{\bx}(t)$ and $V^z_{\bx}(s|t)$ identify the joint
subdistribution $V^z_{\bx}(s,t)$ for $(Y^z_P,Y^z_D)$ given $\bX=\bx$.

\noindent {\bf Assumption 3:} The conditional joint distribution
function of $(\Yd^0, \Yd^1)$ given $\bX=\bx$, $G_{\bx}$, follows a
Gaussian copula model, i.e.,
\begin{equation}
  G_{\bx}(v,w ; \rho) = \Phi_{2, \rho}[ \Phi^{-1}\{ G^0_{\bx}(v) \},
  \Phi^{-1}\{ G^1_{\bx}(w) \} ], 
  \label{eq:assump1}
\end{equation}
where $\Phi$ is a standard normal c.d.f. and $\Phi_{2, \rho}$ is a
bivariate normal c.d.f. with mean 0, marginal variances 1, and
correlation $\rho$.   For fixed $\rho$,  $G_{\bx}$ is identified since
$G^0_{\bx}$ and $G^1_{\bx}$ are identified.
Similar assumptions have been used in the causal mediation literature \citep{daniels2012bayesian}. 


\noindent {\bf Assumption 4:}  Progression time  under treatment $z$   is conditionally independent of death time under treatment $1-z$  given death time under treatment $z$ and covariates $\bX=\bx$, i.e., 
\[
Y_P^{z} \perp Y_D^{1-z}  \; \vline \; Y_D^z , \bX=\bx;   \; \;  z=0,1.
\]


Under Assumptions 1-4, $\tau(\cdot)$ is identified from the distribution of the observed data as follows:
\begin{equation}
\tau(u) = \frac{ \int_{\bx}  \int_{s < u} \int_{v \geq u} \int_{t \geq u} dV^1_{\bx}(s|t) dG_{\bx}(v,t) dK(\bx)  }{ \int_{\bx}  \int_{s < u} \int_{v \geq u} \int_{t \geq u}  dV^0_{\bx}(s|t) dG_{\bx}(v,t) dK(\bx)},
\label{eq:tau}
\end{equation}
where $K(\bx)$ is the empirical distribution of $\bX$.

\section{Bayesian Regression Model}
\label{sec:model}

 In this section, we propose a Bayesian nonparametric  survival regression model on the unknown conditional (on $\bX=\bx$) distribution of $(Y^z_P,Y^z_D)$.  
However, any alternative Bayesian survival regression models could be implemented \citep{hanson2002modeling, gelfand2003bayesian, zhou2017unified, sparapani2016nonparametric}; 
however the first three are restrictive in how covariates are entered and the fourth one is semi-parametric.

%

\subsection{Dependent Dirichlet Process - Gaussian Process Prior}

We start with a review of the Dirichlet process as a
prior for an unknown distribution and step by step extend it to
 the Dependent Dirichlet Process - Gaussian Process prior. 

The Dirichlet process
(DP) prior has been widely used as a prior model for a random
unknown probability distribution.  
We write $H\sim \DP(\alpha, H_0)$
if a random distribution  $H$ of a $J$-dimensional random vector $\bV$  follows a DP prior, 
where $\alpha$ is known as the total mass
parameter and $H_0$ is known as the base measure. 
\cite{sethuraman1991constructive} provides
a constructive definition of a DP, where  
$dH(\bv)=\sum_{h=1}^{\infty}w_h \delta_{\btheta_h}(\bv)$,  $w_h= \nu_h\prod_{l<h}(1-\nu_l)$,
$\btheta_h = \{ \theta_{h1}, \ldots, \theta_{hJ} \} \stackrel{i.i.d.}{\sim} H_0$ and $\nu_h \stackrel{i.i.d.}{\sim} \mathrm{Be}(1, \alpha)$.
In many applications, the
discrete nature of $H$ is not appropriate. A DP mixture model extends
the DP model by replacing each point mass $\delta_{\btheta_h}(\cdot)$ with a
continuous kernel.  For example, a DP mixture of normals takes the form: $dH(\bv)=\sum_h w_h \phi \left( \bv; \btheta_h, \bSigma \right) d \bv$, where $\phi(\cdot;\bmu,\bS)$ is the density function of a multivariate normal random vector with mean vector $\bmu$ and variance-covariance matrix $\bS$.

To introduce a prior on the conditional (on covariates $\bX=\bx$) distribution ($H_{\bx}$) of $\bV$,
the DP mixture model has been extended to a dependent DP (DDP) by replacing  $\btheta_h$ in each term
with $\btheta_h(\bx) = \{\theta_{h1}(\bx), \ldots, \theta_{hJ}(\bx)\}$, which is a multivariate stochastic process indexed by $\bx$.  A DDP mixture of normals takes the form: 
\begin{equation}
\label{dHx}
dH_{\bx}(\bv)=\sum_h w_h \phi \left( \bv; \btheta_h(\bx), \bSigma \right) d \bv.
\end{equation}  
To complete the prior specification, we need to posit a stochastic process prior for $\{\btheta_{h}(\bx): \bx\}$.
A common specification are independent Gaussian process (GP) priors
\citep{maceachern1999dependent} on $\{\theta_{hj}(\bx): \bx\}$.  A GP
prior is specified such that for all $L \geq 1$ and
($\bx_1,\ldots,\bx_L)$,  
the distribution of $(\theta_{hj}(\bx_1), \dots, \theta_{hj}(\bx_L))$ follows 
a multivariate normal distribution with mean vector 
$(\mu_{hj}(\bx_1), \dots, \mu_{hj}(\bx_L))$ and $(L \times L)$
covariance matrix where the $(l, l')$ 
entry is $R_{j}(\bx_l, \bx_{l'}$).  We write $\{\theta_{hj}(\bx): \bx\}
\sim \GP(\mu_{hj}(\cdot), R_j(\cdot,\cdot))$.   
For an extensive review of the GP priors, see
\cite{Rasmussen:2006} and \cite{MacKay1999}.  
We model the mean function $\mu_{hj}(\cdot)$ as a linear regression on covariates
$ 
\mu_{hj}(\bx_l; \bbeta_{hj})=\bx_l \bbeta_{hj},
$ 
with covariance process specified as
\begin{equation}
  R_{j}(\bx_l, \bx_{l'}) = \exp \left\{ -\sum_{d=1}^D (x_{l d}-x_{l' d})^2
  \right\} + \delta_{l l'} \epsilon ^2, 
 \label{eq:cov}
\end{equation}
where $D$ is the dimension of the 
covariate vector, $\delta_{l l'}=I(l=l')$ and $\epsilon$ is a small constant (e.g., $\epsilon=0.1$) used to ensure that the covariance function is positive definite. 
To ensure a reasonable covariance structure, continuous covariates should be standardized to have mean 0 and variance 1. More flexible covariance functions can be considered if desired. 
Additional priors are introduced on the $\bbeta_{j}$'s and $\bSigma$, the details of which are discussed in Appendix A1. 
%
We write $\{H_{\bx}\} \sim 
\DDPGP(\alpha, \Sig, \GP(\mu_j(\cdot), R_j(\cdot,\cdot)),
j=1,\ldots,J)$. 

 
\subsection{Application to Semi-competing Risks Data}

Separately for each treatment group $z$, we posit independent DDP-GP's
on the unknown conditional (on $\bX=\bx$) probability measure ($H^z_{\bx}$) of $(Y^z_P,Y^z_D)$.  
Since $V^z_{\bx}(s|t) = H^z_{\bx}(Y_P^z \leq s, Y_P^z \leq Y_D^z | Y^z_D=t)$ ($s \leq t$) and $G^z_{\bx}(t) =  H^z_{\bx}(Y^z_D \leq t)$,
the prior on $H^z_{\bx}$ induces priors on $V^z_{\bx}(s|t)$ and $G^z_{\bx}(t)$ (identified under Assumptions 1 and 2) 
and together with the Gaussian copula for $G_{\bx}$ implies
a prior on the estimand $\tau(\cdot)$.  The prior on $H^z_{\bx}$
also induces priors on non-identified quantities which have no impact on our 
analysis. More specifics about our prior are presented in Appendix A1.

Before transitioning to the posterior sampling algorithm, note that the relevant portion of the observed data likelihood for individual $i$, with data $O_i = (T_{1i}, T_{2i}, \delta_i, \xi_i,
Z_i,\bX_i)$ is
\begin{eqnarray*}
L(O_i)  &= & \left\{  dV^{Z_i}_{\bX_i}(T_{1i}|T_{2i}) dG^{Z_i}_{\bX_i} (T_{2i})\right\}^{\delta_i \xi_i  } \left\{  \int_{t > T_{2i}}  dV^{Z_i}_{\bX_i}(T_{1i}|t) dG^{Z_i}_{\bX_i} (t) \right\}^{\delta_i (1-\xi_i) }  \times \\
 & & \left\{ \left( 1- V^{Z_i}_{\bX_i}(T_{2i}|T_{2i}) \right) dG^{Z_i}_{\bX_i} (T_{2i}) \right\}^{(1-\delta_i) \xi_i }  
\left\{ \int_{t > T_{2i}}  \left(   1- V^{Z_i}_{\bX_i}(T_{2i}|t) \right) dG^{Z_i}_{\bX_i} (t)  \right\}^{(1-\delta_i) (1-\xi_i) } \\  
&= & \left\{  dH^{Z_i}_{\bX_i}(T_{1i},T_{2i}) \right\}^{\delta_i \xi_i  } \left\{  \int_{t > T_{2i}}  dH^{Z_i}_{\bX_i}(T_{1i},t)  \right\}^{\delta_i (1-\xi_i) }  \left\{  \int_{s> T_{1i}} dH^{Z_i}_{\bX_i} (s,T_{2i}) \right\}^{(1-\delta_i) \xi_i }  \times \\
& & \left\{ \int_{t > T_{2i}} \int_{s > T_{2i}} dH^{Z_i}_{\bX_i}(s,t)  \right\}^{(1-\delta_i) (1-\xi_i) }.  \end{eqnarray*}
We include the second equality because it allows us to see that, using data augmentation to replace the integrals, the joint full data likelihood is $\prod_{i=1}^n dH^{Z_i}_{\bX_i}(Y_{Pi},Y_{Di})$.  This will allow us to use
  existing posterior simulation techniques for DDP-GP models.

\subsection{Posterior Simulation}
\label{sec:mcmc}

The details of the MCMC algorithm are presented in Appendix A2.  Here
we focus on individuals assigned to treatment $z$ and suppress the
dependence of the notation on $z$. As noted above, the MCMC
implementation is based on the full data likelihood.  While $dH_{\bX_i}$ is an
infinite mixture of normals, we approximate it by a finite mixture with $K < \infty$ components.  
This finite mixture model 
for $(Y_{Pi},Y_{Di})$ can be
replaced by a hierarchical model where (1) $\gamma_i$ is a latent
variable that selects mixture component $h$ ($h=1,\ldots, K$) with
probability $w_h$ (properly normalized to handle the finite number of mixture components) and (2) given $\gamma_i$, the pair $(Y_{Pi},Y_{Di})$ follows
a multivariate normal distribution with mean $\btheta_{\gamma_i}(\bX_i)$ and
variance $\bSigma$.
%

Posterior simulation is based on this hierarchical model characterization.  Importantly, all of the full conditionals in the MCMC algorithm have a closed form representation.    Details of the Markov chain Monte Carlo posterior simulation can be
found in Appendix A2.


%

\section{Simulation Studies}
\label{sec:simu}
\subsection{Simulation Setup}

We considered three simulation scenarios to evaluate the performance of our
proposed approach with 500 repeated simulations for each scenario. 
We generated $Z \sim \Bern(0.5)$.  Independently of $Z$, we generated two independent covariates $X_1$ and $X_2$, where $X_1$ followed a truncated normal distribution with mean 4.5, variance 1 and truncation interval $(2,7.5)$ and $X_2 \sim Bern (0.4)$.
For the first two simulation scenarios, we simulated progression time and death time on the log scale as follows: 
\begin{eqnarray*}
   Y^z_{P} & = & 1.5z +0.6X_{1}+2X_{2}+\epsilon,\\
   \Yd^z  & =  & 4z +0.3X_{1}+X_{2}+\nu.
\end{eqnarray*}

In \underline{Scenario 1}, we assumed $(\epsilon,\nu)$ followed a  bivariate normal distribution with mean $(0, 1.5)'$, marginal variances $S_{11}=S_{22}=1$, and correlation $S_{12}=0.75.$
 In \underline{Scenario 2}, we assumed $(\epsilon, \nu)$ to be a scaled multivariate $t$ distribution with degree of freedom $\nu=3$, mean $(0,1.5)'$, marginal
variance $S_{11}=S_{22}=1$, and 
correlation $S_{12}=0.75$. 
\underline{Scenario 3}  explored performance under a nonlinear covariate effect specification on progression and death times. We generated $Y_{P}^z  =  1.5z+0.6X_{i}+2X_{2}+\epsilon$ and 
   $\Yd^z =  4z +0.3X_{1}+X_{2}+0.5\sqrt{X_{1}}+\nu$, with $(\epsilon, \nu)$ following the same distribution as in Scenario 1. 

In all scenarios, the censoring time $C^z$ on the log scale was generated independently according to a $\Unif(8, 10)$ distribution. For the joint distribution of $Y_D^0$ and $Y_D^1$ in \eqref{eq:assump1}, we set $\rho=\rho^o=0.5$ in the Gaussian copula as the truth. We  
 generated $(Z_i,X_{1i},X_{2i},\Ypi, \Ydi,C_i)$ for $n=500$ independent patients and then coarsened to $(Z_i,X_{1i},X_{2i},T_{i1}, T_{i2}, \delta_i, \xi_i)$.

To explore sensitivity of $\tau(u)$ with respect to $\rho$, we conducted inference for $\tau(u)$ under several values of $\rho=0.2, 0.5, 0.8$. For all three scenarios, we specified hyperparameters as described in Appendix A1. 

For comparative purposes, we implemented a naive Bayesian (Naive) model by assuming that the conditional probability measure ($H_{\bx}^z)$ of $(Y_P^z, Y_D^z)$ follows a multivariate normal distribution with mean $(\bx' \bbeta_P^z, \bx' \bbeta_D^z)$ and variance-covariance matrix $\bS^z$, with conjugate multivariate normal priors on $\bbeta_P^z$ and $\bbeta_D^z$ and an inverse Wishart prior on $\bS^z$ (i.e., $\bbeta_P^z\sim \mathrm{MN}({\bm 0}, \tau_P^2{\bm I})$, $\bbeta_P^z\sim \mathrm{MN}({\bm 0}, \tau_D^2{\bm I})$, and $\bS^z\sim \mathrm{Inverse \ Wishart}(\nu, \Psi)$, $z=0, 1$). 
 
For each analysis, we ran 5,000 MCMC iterations with an initial burn-in of 2,000 iterations and a thinning factor of 10. The convergence diagnostics using the R package \texttt{coda} show no evidence of practical convergence problems.

\subsection{Simulation Results} 

\begin{figure}
\centering
\begin{tabular}{ccc}
Scenario 1& Scenario 2 & Scenario 3 \\ \\
\multicolumn{3}{c}{Treatment 0} \\
\includegraphics[scale=0.3]{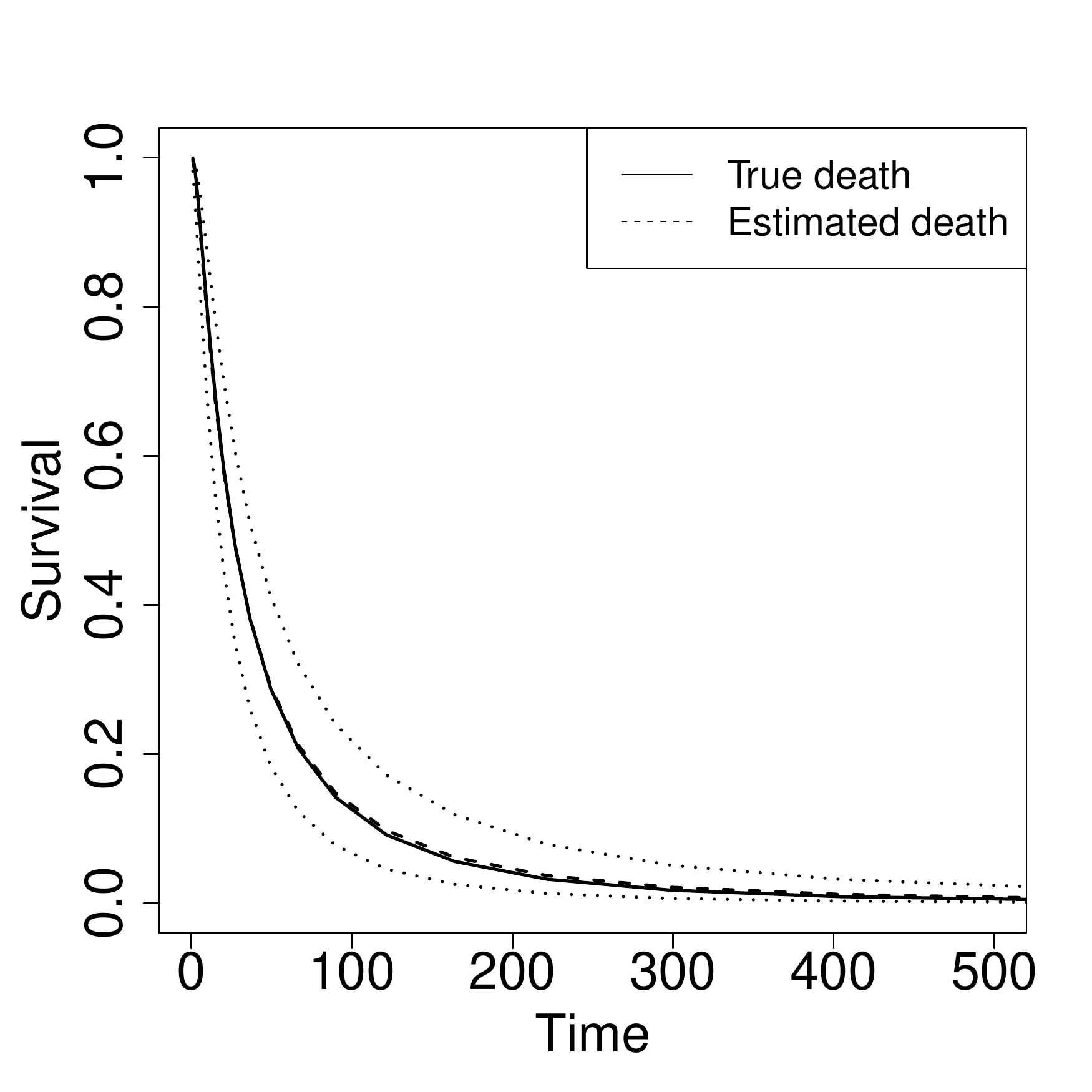} & \includegraphics[scale=0.3]{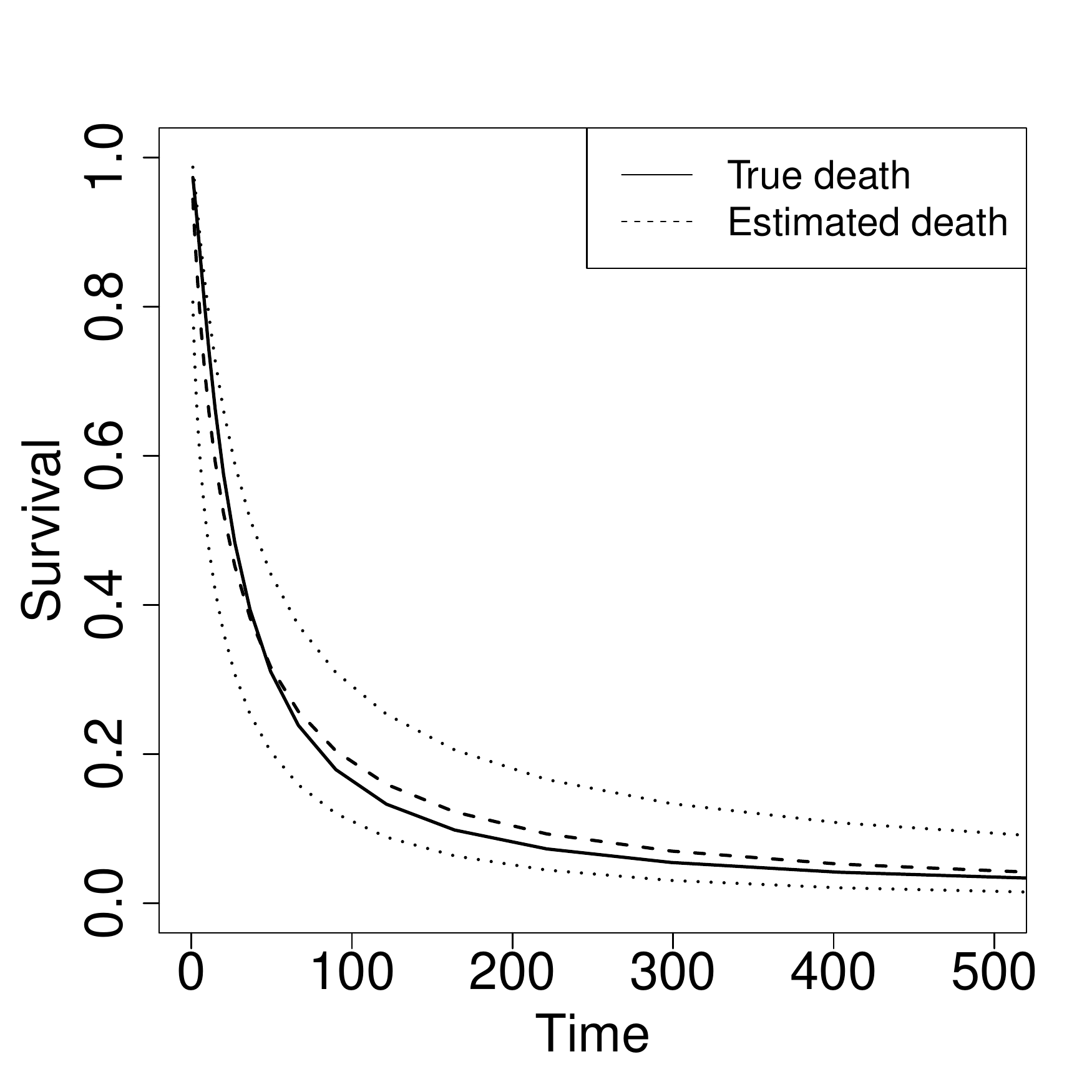} &  \includegraphics[scale=0.3]{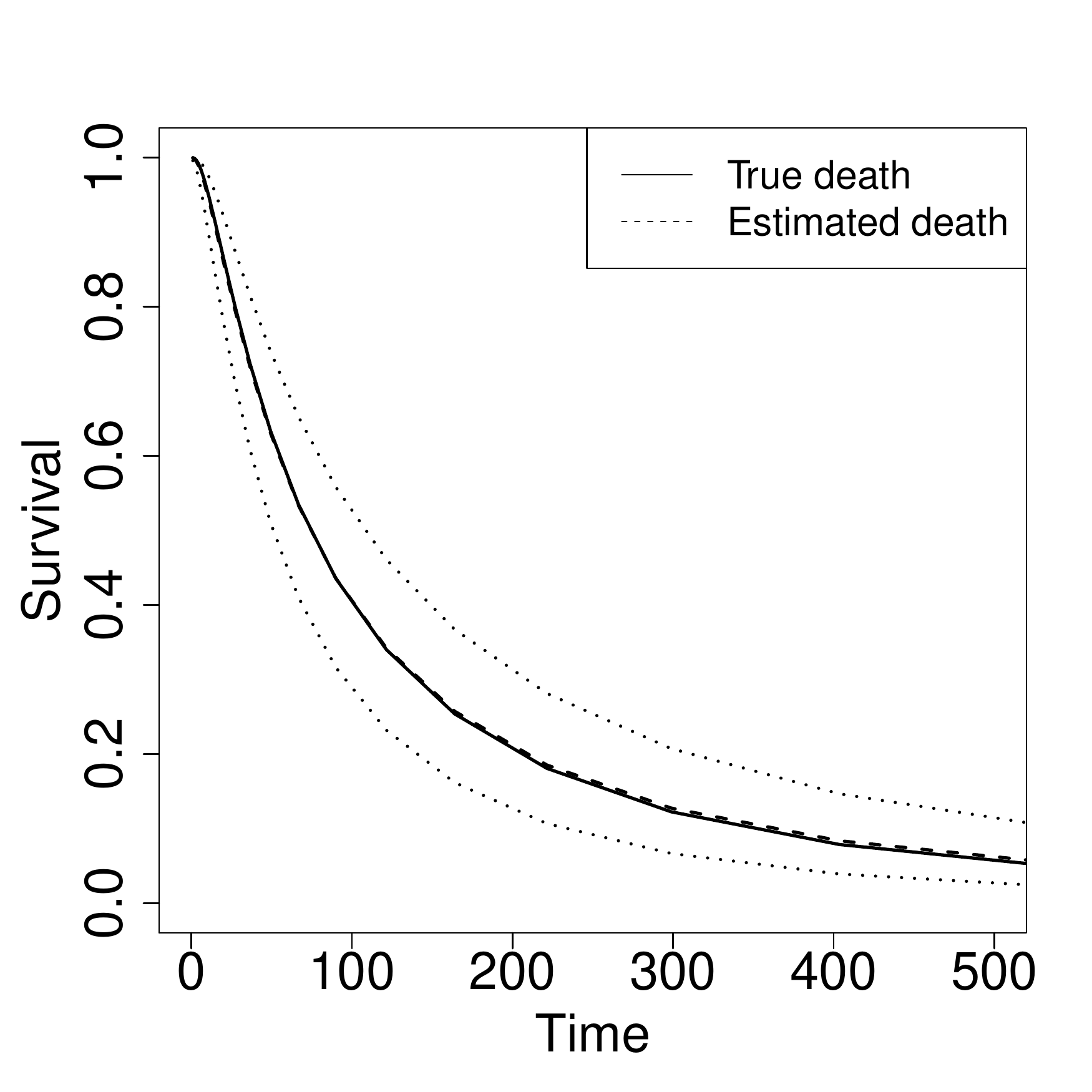}\\
\multicolumn{3}{c}{Treatment 1} \\
\includegraphics[scale=0.3]{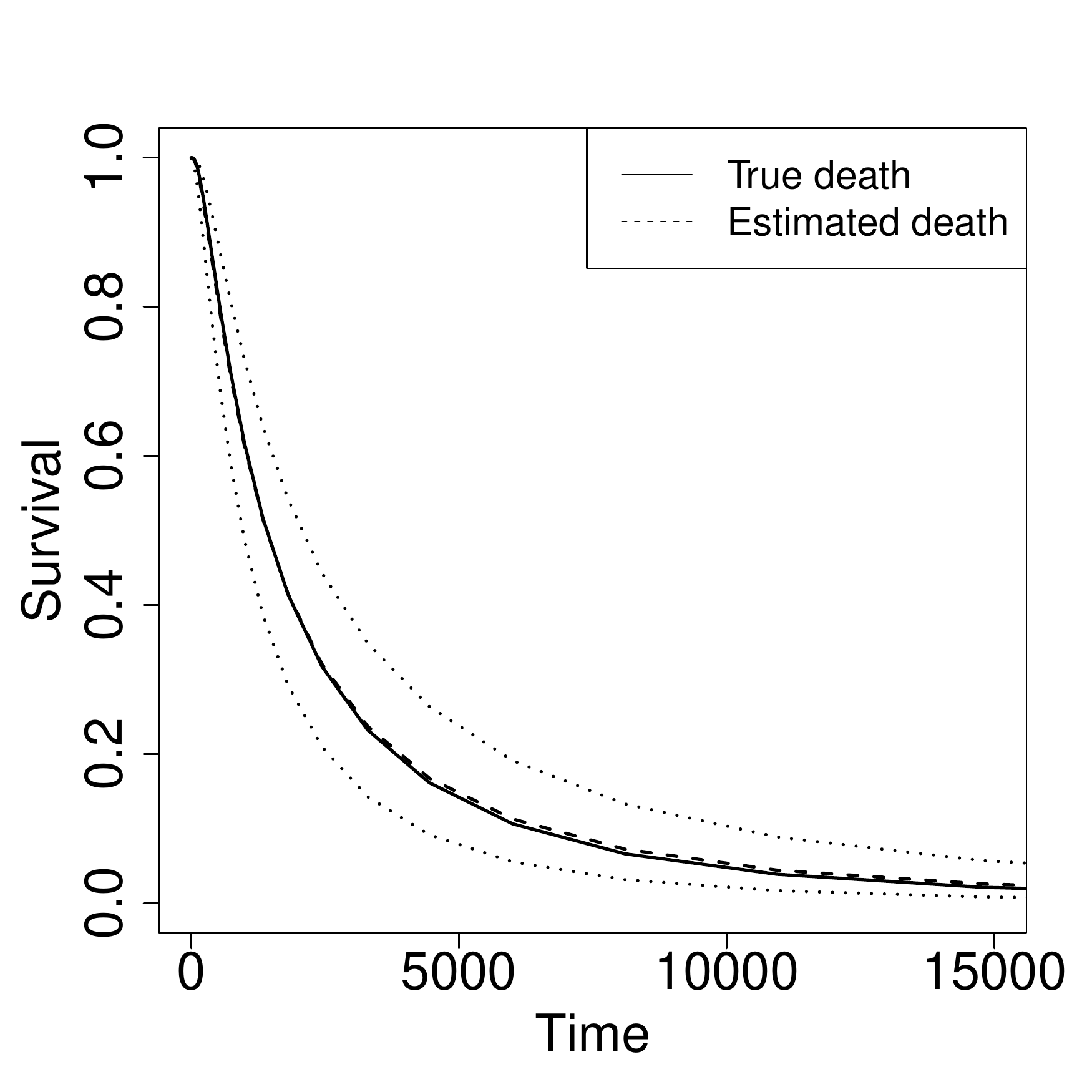} & \includegraphics[scale=0.3]{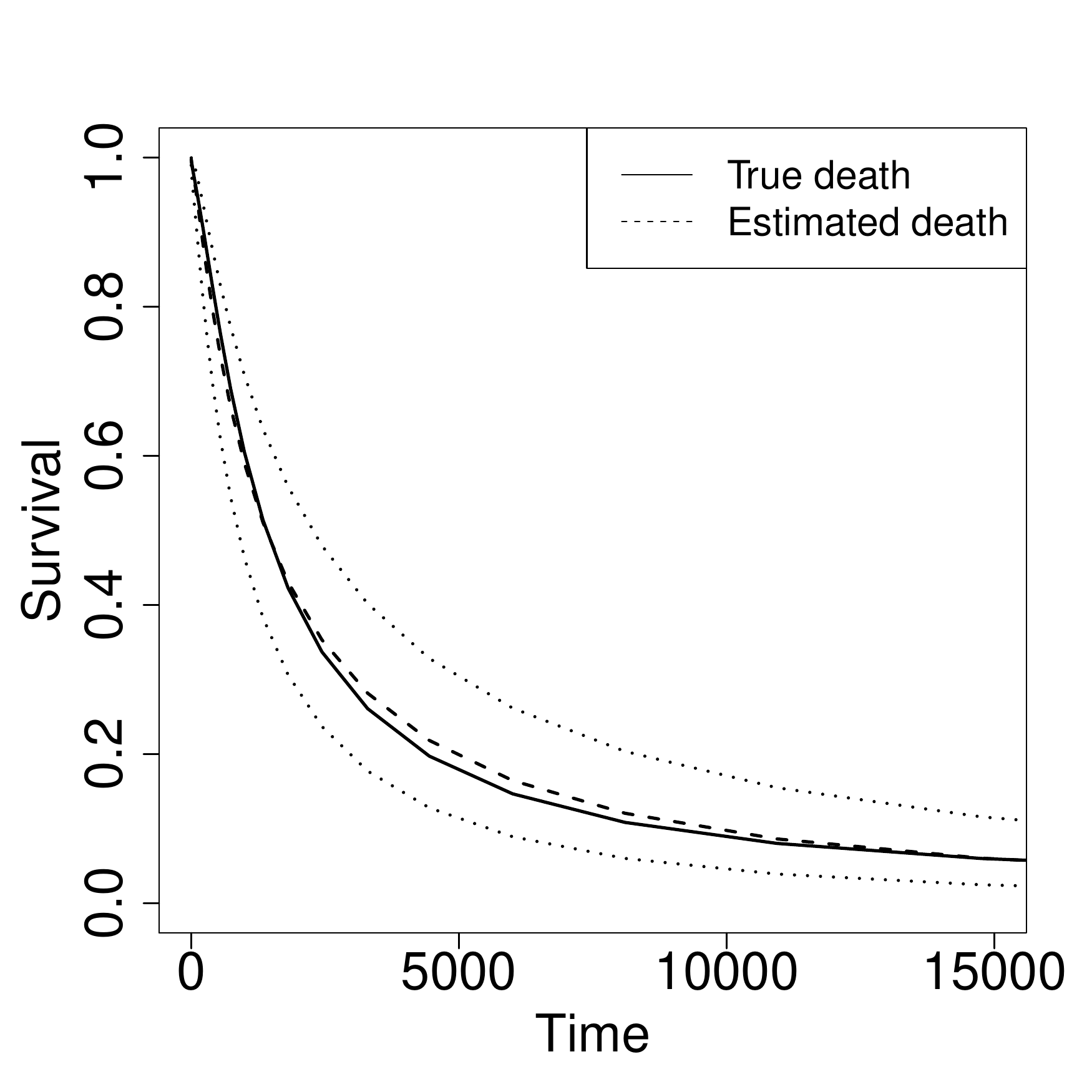} &  \includegraphics[scale=0.3]{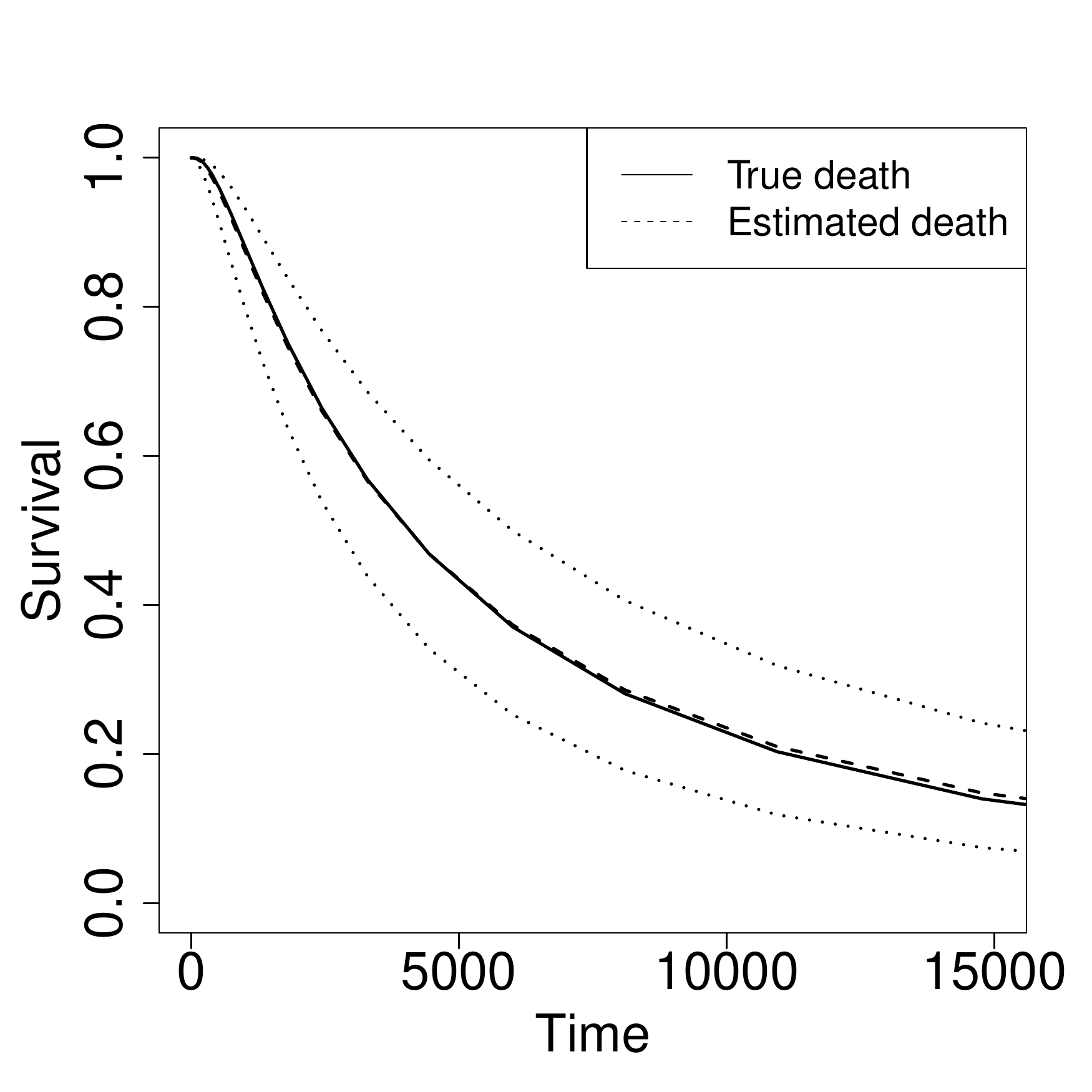}  \\
\end{tabular}
  \vspace{.6cm}
\caption{For each simulation scenario and by treatment group (first and second rows refer to treatments 0 and 1, respectively), the true survival functions (solid line), the posterior mean survival functions averaged over simulated datasets (dashed line), and 95\% point-wise credible intervals (computed using quantiles) averaged over simulated datasets (dotted lines). Survival times are on the original scale (days). }
\label{fig:marginal}
\end{figure}

We first report on the performance in terms of recovering the true treatment-specific marginal survival functions for time to death. For the BNP approach, Figure \ref{fig:marginal} shows, for each of the three simulation scenarios and by treatment group (first and second rows refer to treatments 0 and 1, respectively), the true survival functions (solid line), the posterior mean survival functions averaged over simulated datasets (dashed line), and 95\% point-wise credible intervals (computed using quantiles) averaged over simulated datasets (dotted lines) on the original time scale (days). As another metric of performance, we computed, for each simulated dataset, the root mean squared error (RMSE) taken as  the square root of the average of the squared errors at 34 equally-spaced grid points in  log-scaled time interval $(0, 10)$.  For each scenario, Table \ref{table:simu} summarizes the mean and standard deviation of RMSE across the 500 simulated datasets.  Both Figure \ref{fig:marginal} and 
Table \ref{table:simu} show that our proposed BNP procedure performs well, for each of the three scenarios, in terms of recovering the true survival function.

Table \ref{table:simu} also shows the mean and standard deviation of RMSE for the Naive model. In scenario 1, the model matches the simulation true model, thereby yielding comparable results as the proposed BNP model. In contrast,  the Naive model performs worse than the BNP model in scenario 2 when the fitted model does not match the simulation truth. In scenario 3, the BNP model performs slightly better than the Naive model. Overall, the proposed BNP model is more robust compared to the Naive model.

Evaluation of $\tau(\cdot)$ requires evaluation of $G_{\bx}^z$ as the second
marginal under $H_{\bx}^z$. Expression \eqref{eq:tau} allows us now to estimate $\tau(u)$.
Both the numerator and denominator can be
 evaluated as functionals of the currently imputed random
probability measure $H_{\bx}^z(\Yp^z, \Yd^z)$  of 
time to log progression $\Yp^z$ and time to log death $\Yd^z$ under treatment $z$,  marginalizing with respect to the empirical distribution of $\bx$. Each iteration of the posterior 	MCMC simulation evaluates a point-wise estimate and we estimate the posterior mean of $\tau(u)$ as 
$\hat{\tau}(u)= \mathrm{Mean}\left\{\tau(u) \mid data\right\}$ across iterations. 
We also report the mean RMSE in estimating the $\hat{\tau}(u)$ by averaging over 500 repeated simulations under the proposed BNP model and the Naive model. Table  \ref{table:simuh} summarizes the results.

\begin{figure}
\centering
\begin{tabular}{ccc}
\multicolumn{3}{c}{Scenario 1} \\ 
  \includegraphics[scale=0.28]{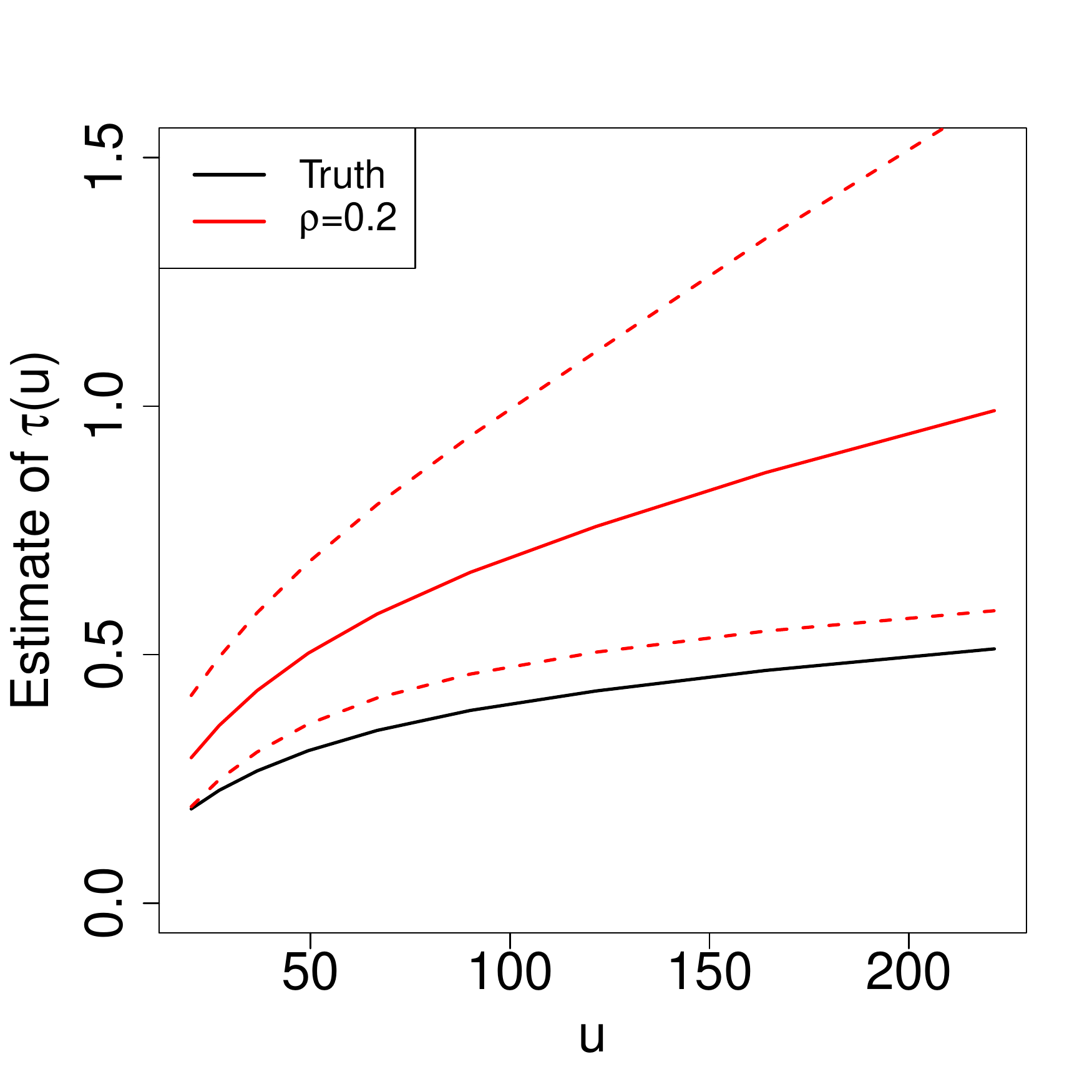} &  \includegraphics[scale=0.28]{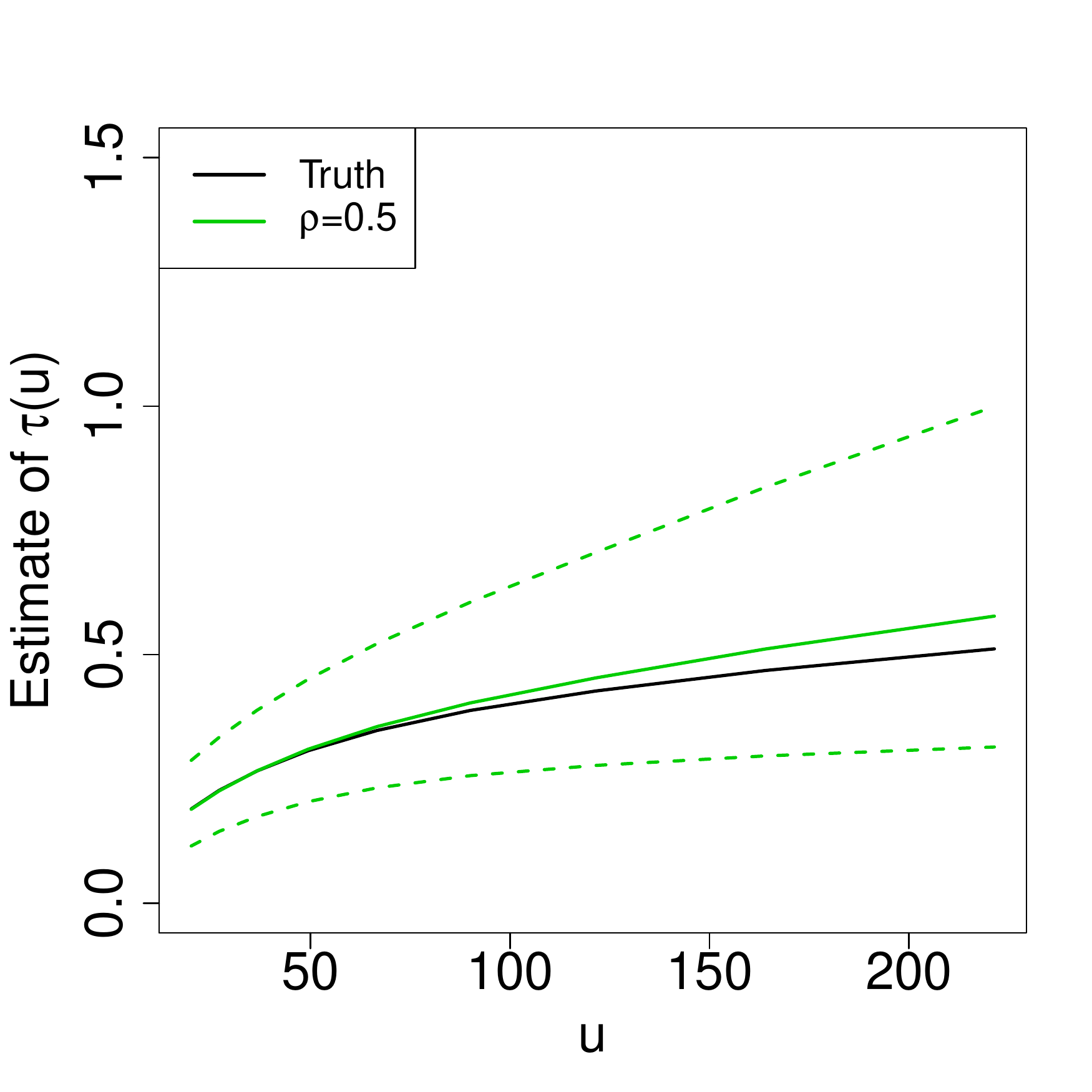} &   \includegraphics[scale=0.28]{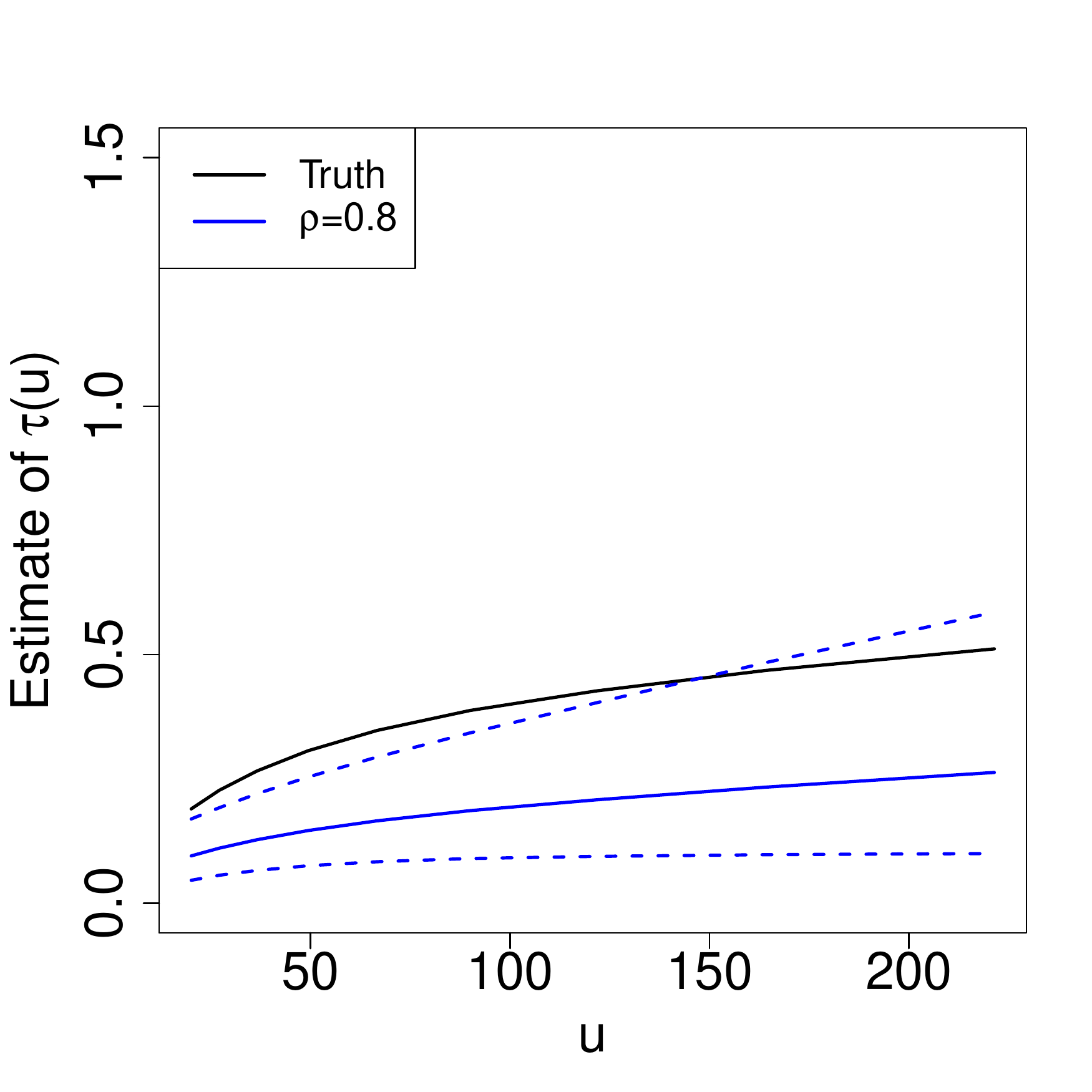} \\
\multicolumn{3}{c}{Scenario 2} \\ 
  \includegraphics[scale=0.28]{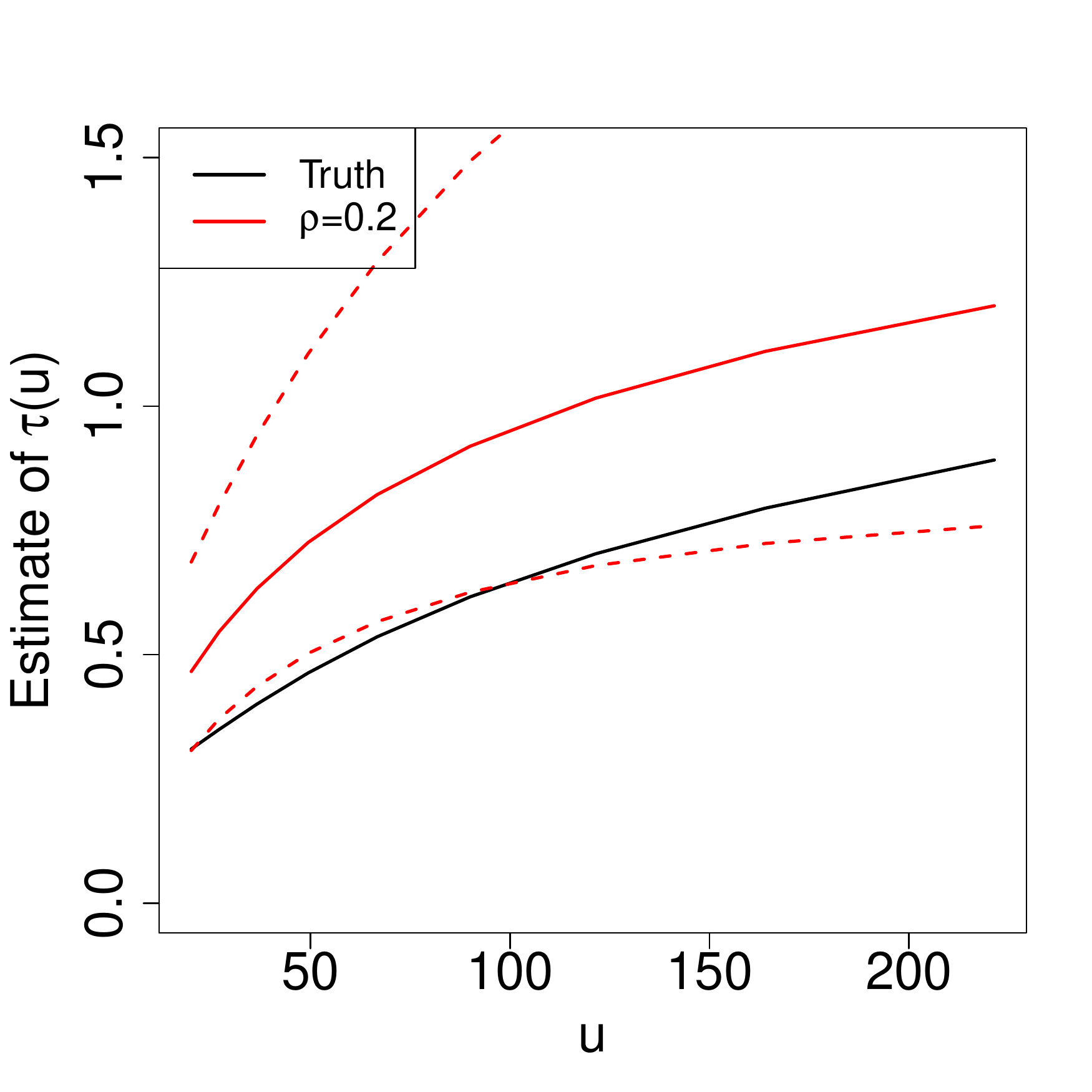} &  \includegraphics[scale=0.28]{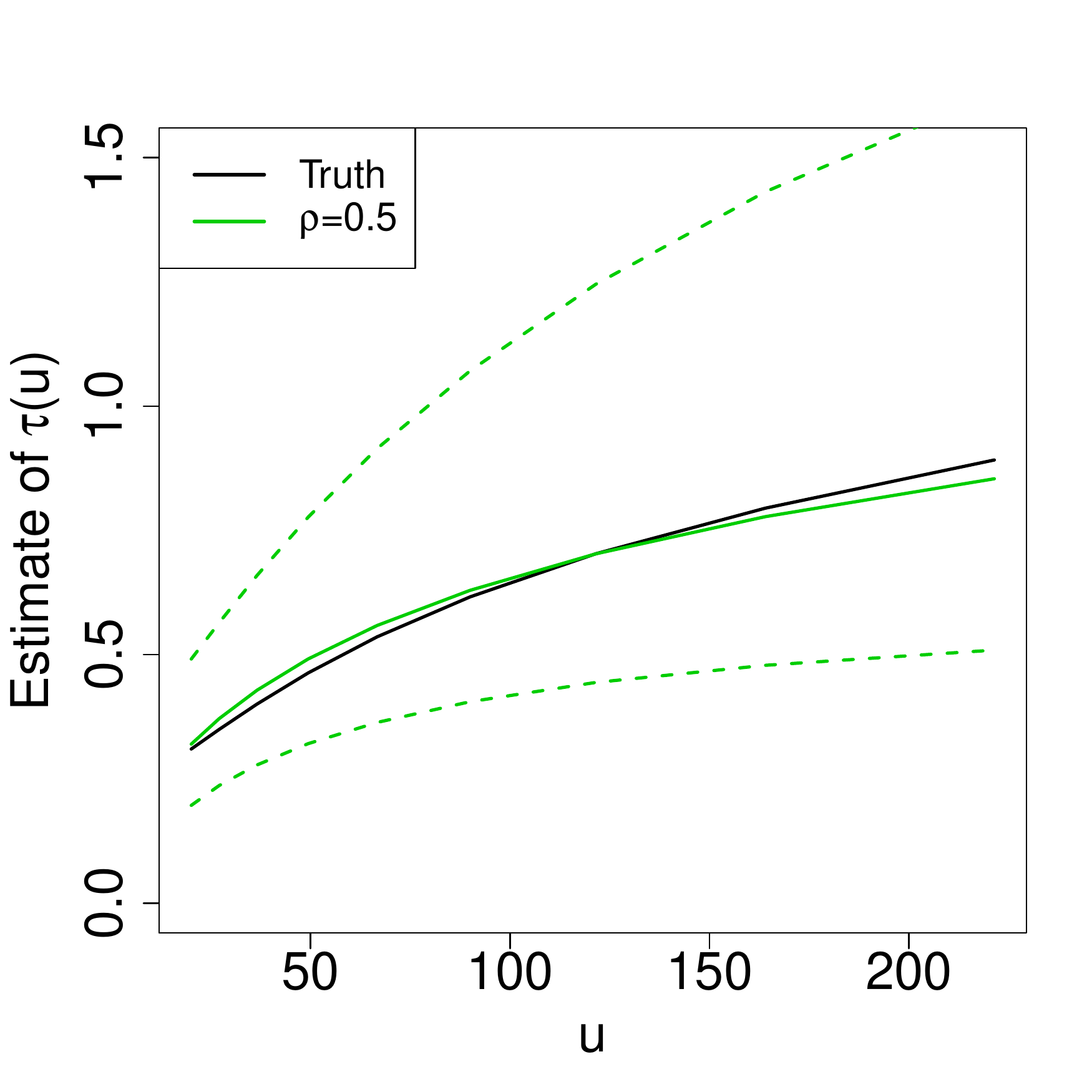} &   \includegraphics[scale=0.28]{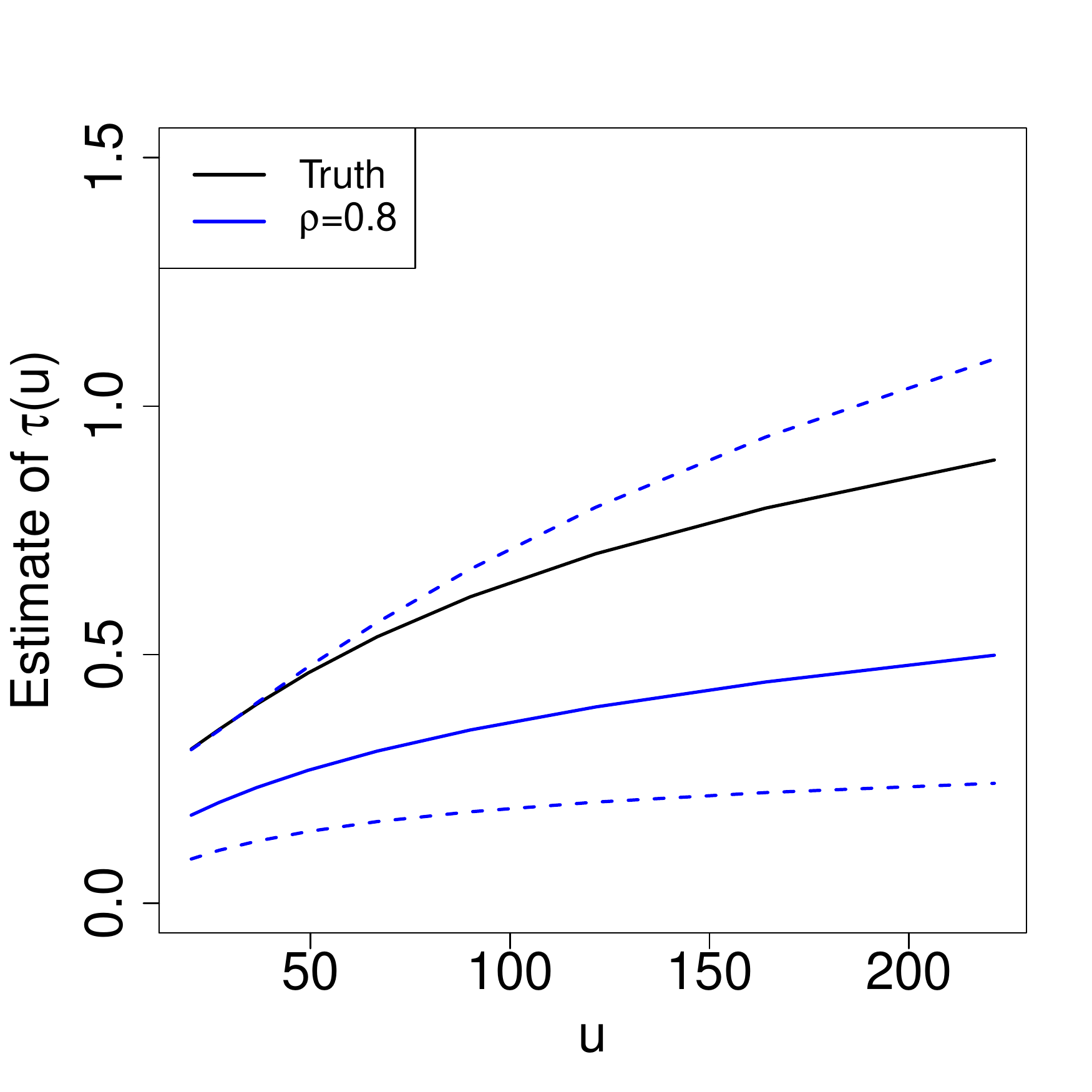} \\
\multicolumn{3}{c}{Scenario 3} \\ 
  \includegraphics[scale=0.28]{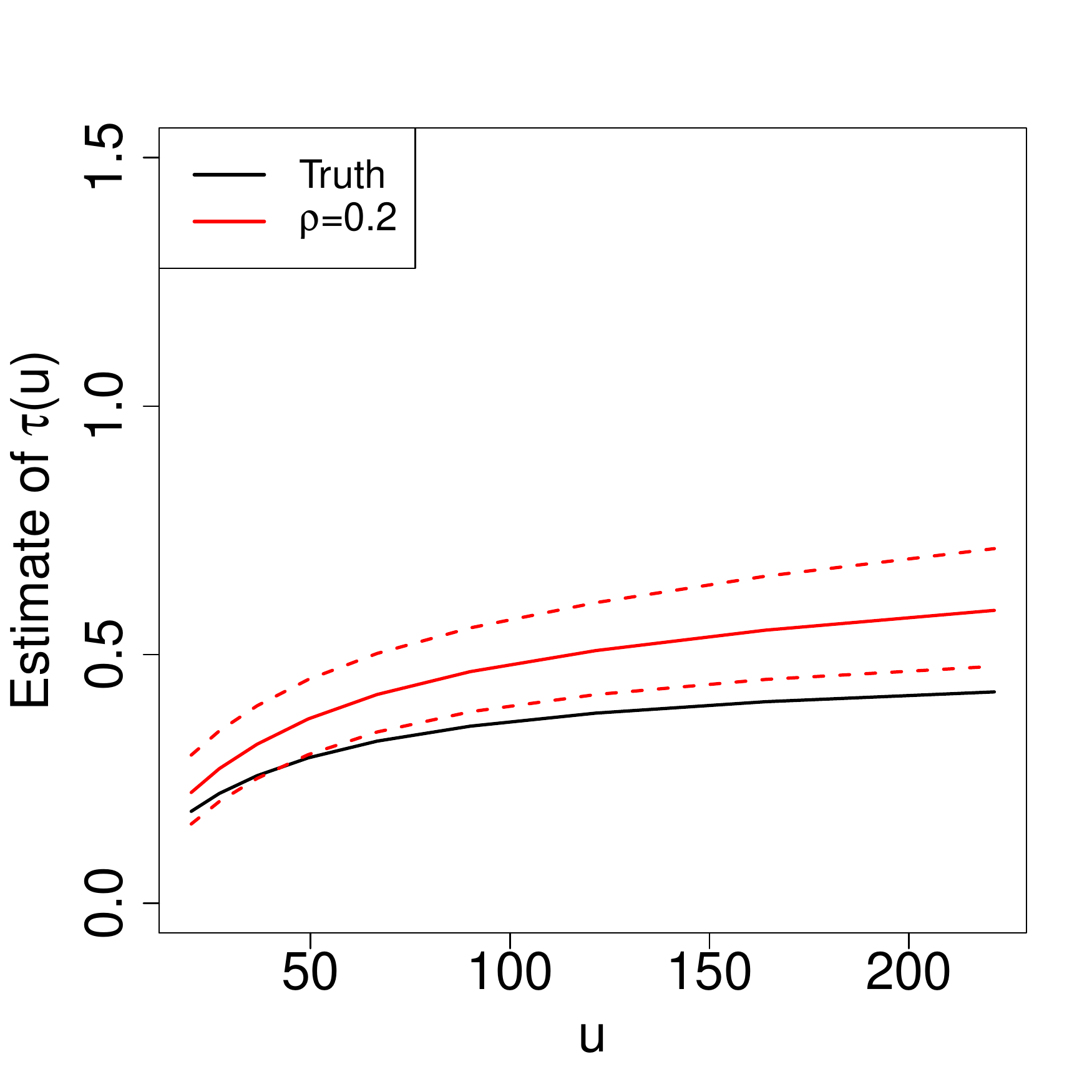} &  \includegraphics[scale=0.28]{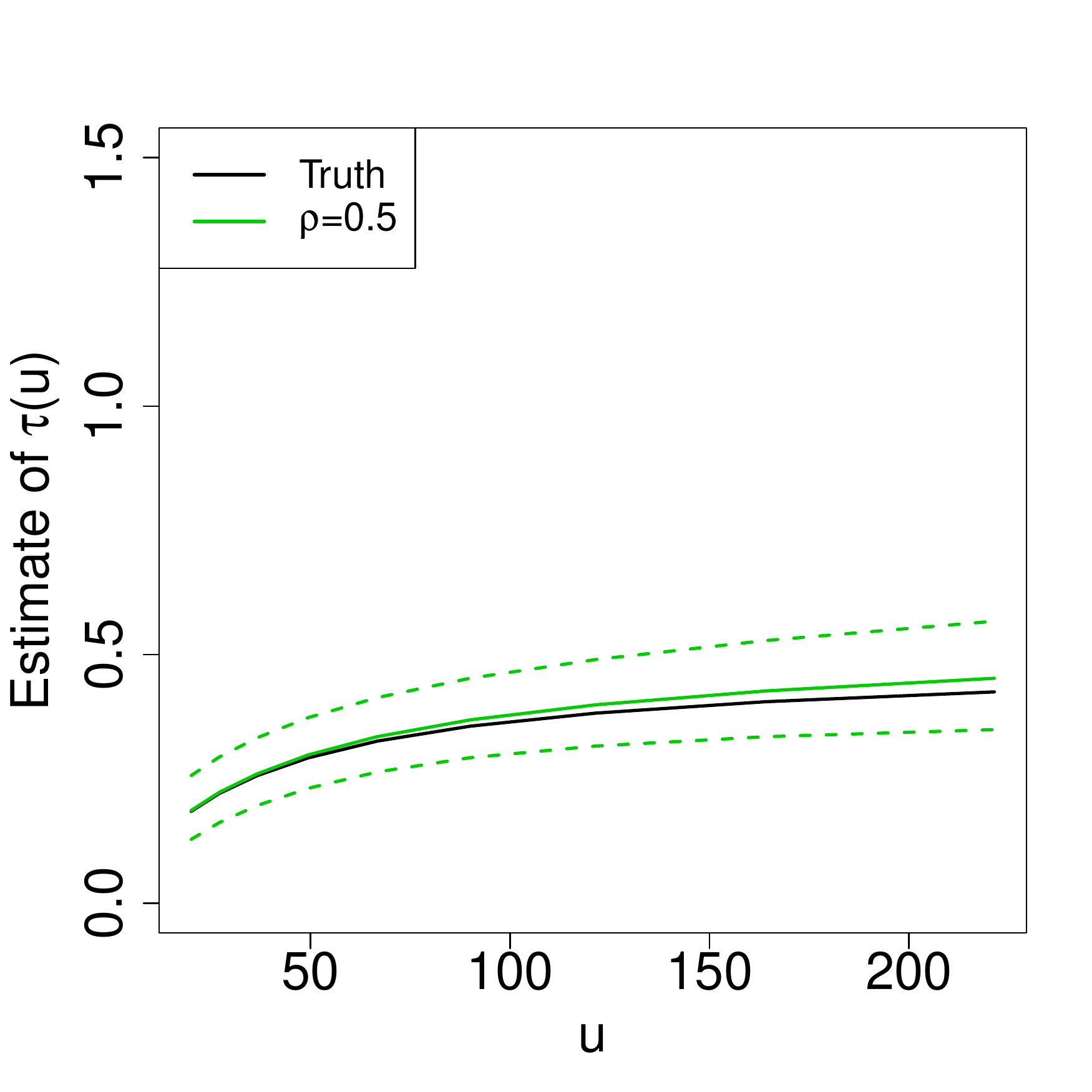} &   \includegraphics[scale=0.28]{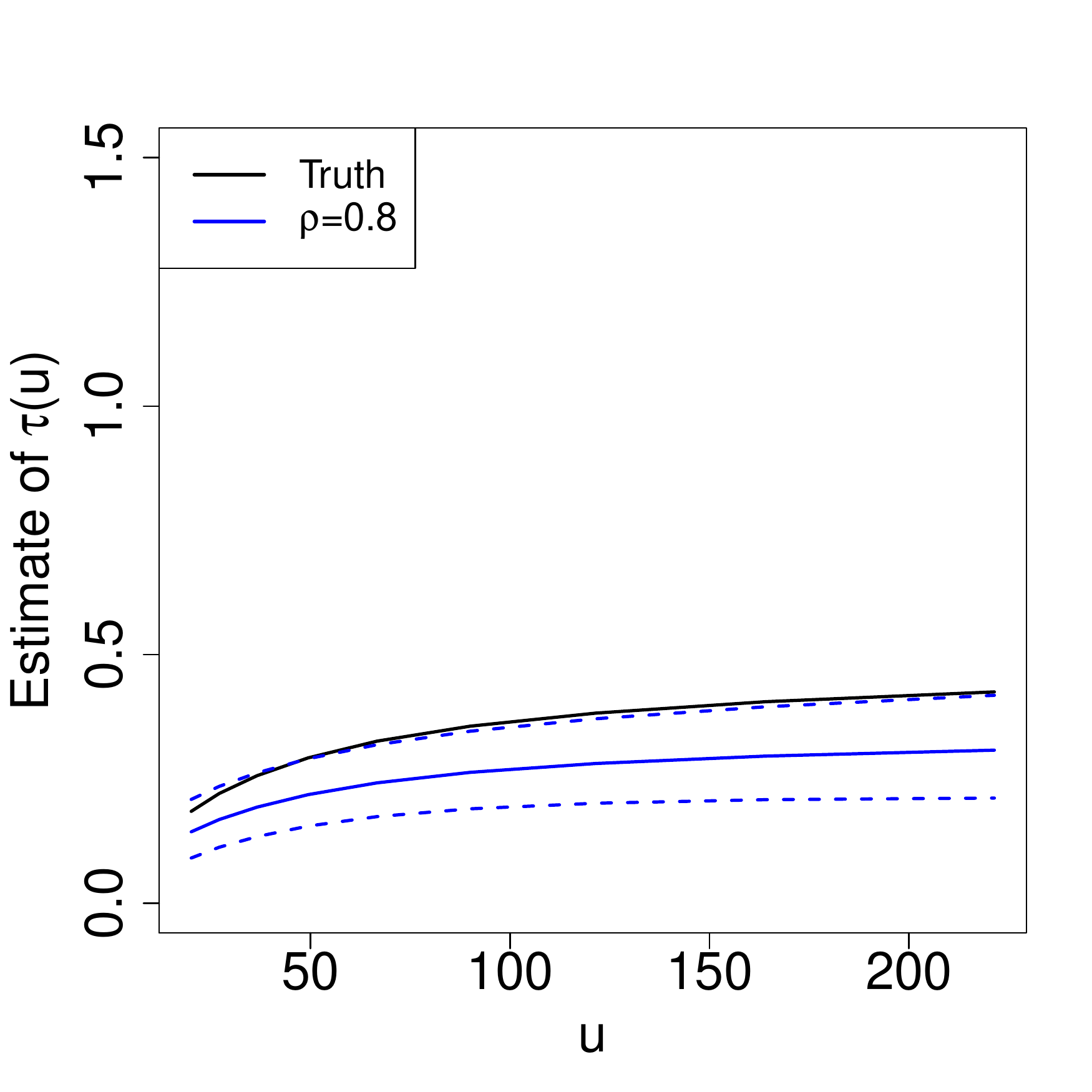} \\
\end{tabular}
  \vspace{.6cm}
\caption{The posterior estimates of $\tau(u)$ versus $u$ on the original scale (days) for the three scenarios using $\rho=0.2, 0.5, 0.8$, respectively.  The black  lines represent the simulation truth using $\rho^o=0.5$. The dashed lines represent 95\% point-wise credible intervals (computed using quantiles) averaged over simulated datasets. }
\label{fig:h}
\end{figure}

Figure \ref{fig:h} shows $\hat{\tau}(u)$ versus $u$ in the three scenarios, respectively, using $\rho=0.2, 0.5$ and $0.8$. 
As shown in Figure \ref{fig:h}, 
in all three scenarios, 
when $\rho=\rho^o=0.5$, the estimates under the proposed BNP model reliably recover the simulated 
 true $\tau(u)$ and avoid the excessive bias seen with  other $\rho$ values. This agrees with the results reported in Table \ref{table:simuh} that  $\rho=0.5$ always yields the smallest mean RMSE in all three scenarios. Furthermore, when $\rho=0.5$, the proposed BNP model has smaller mean RMSE compared to the Naive model. When $\rho=0.2$ or $0.8$, the BNP model performs better or comparable to the Naive model in terms of providing smaller mean RMSE and variability of RMSE across simulations. 

\section{Brain Tumor Data Analysis}
\label{sec:brain}

\begin{figure}
  \begin{tabular}{cc}
  \includegraphics[width=.5\textwidth]{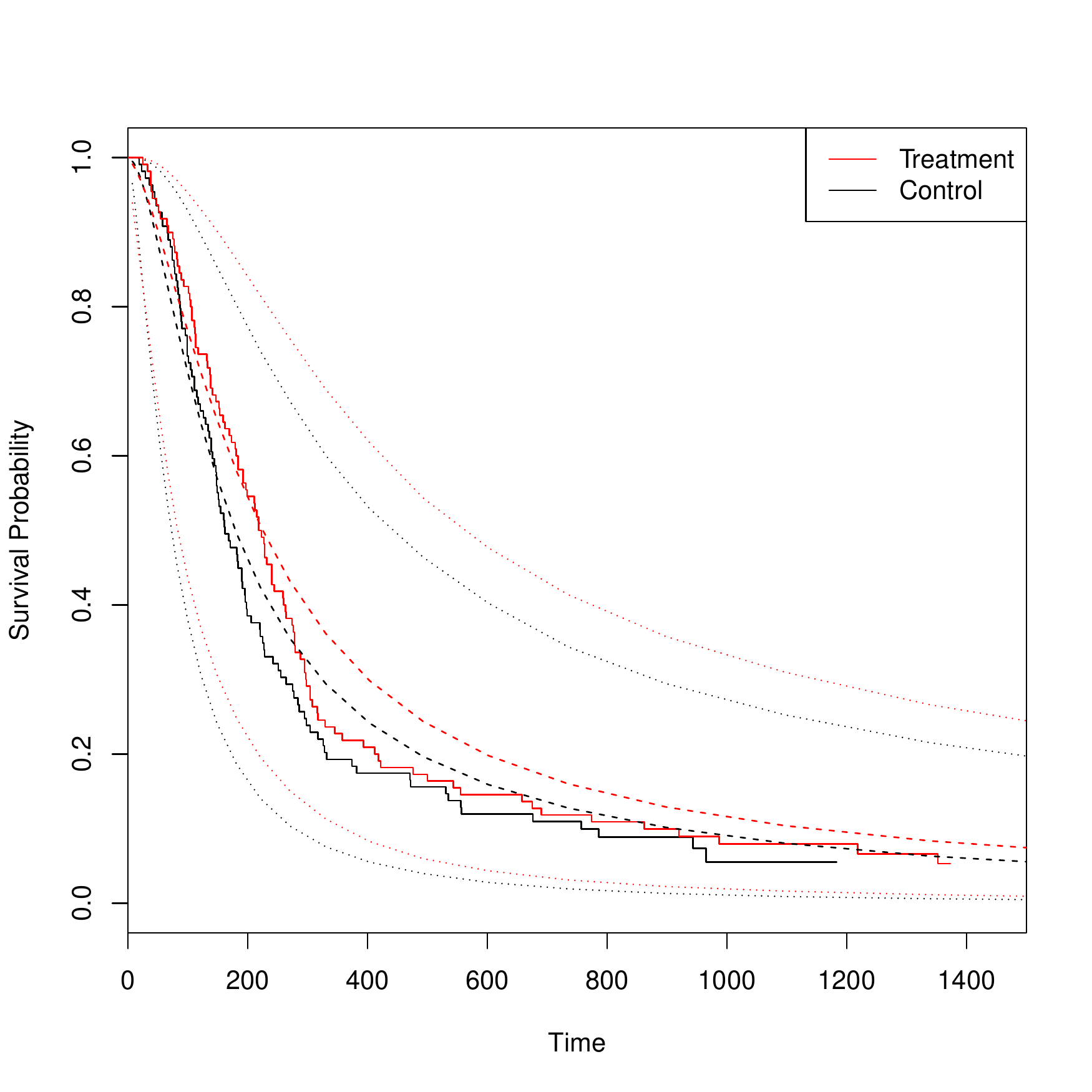}
  & \includegraphics[width=.5\textwidth]{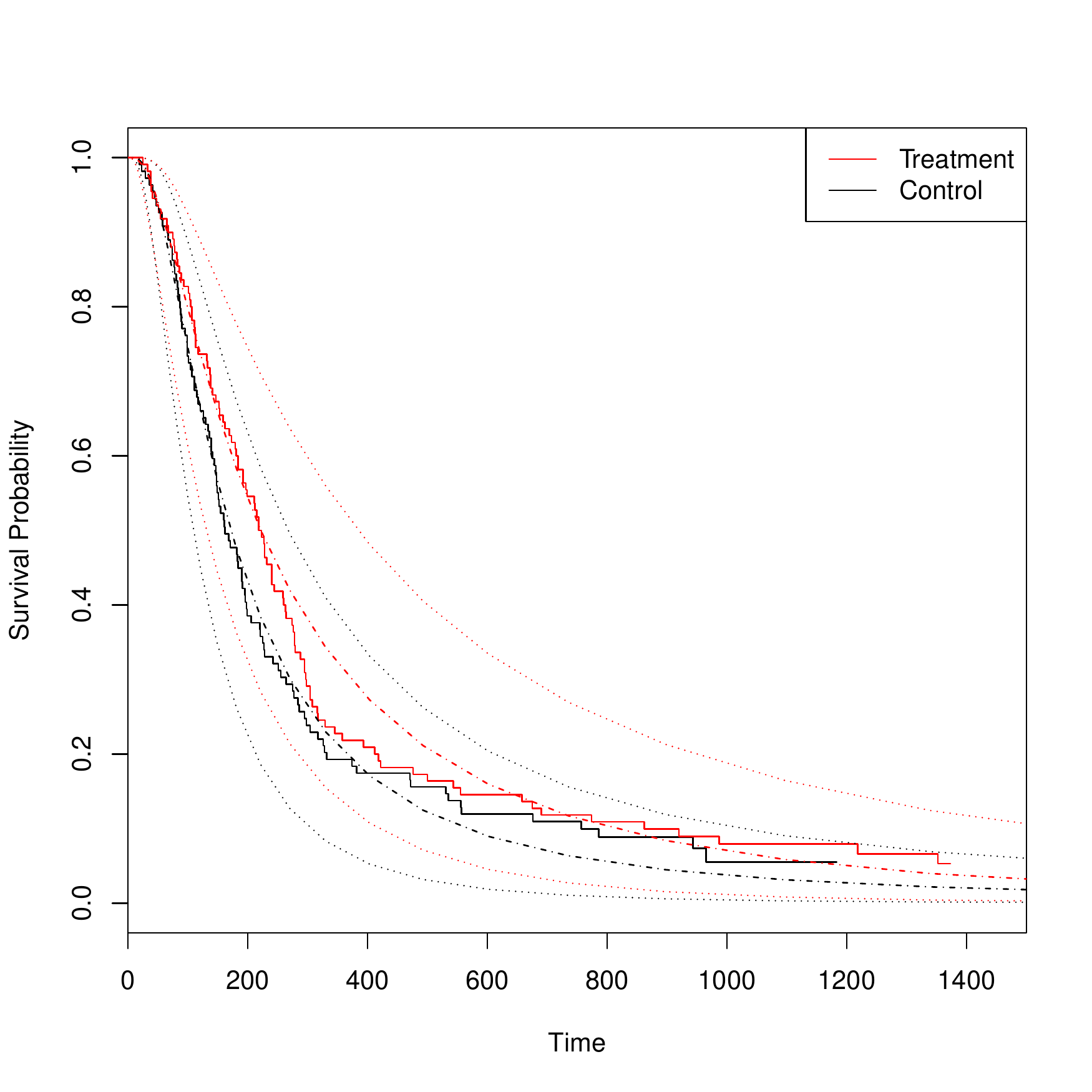}\\ 
(a) BNP &(b) Naive\\
\end{tabular}
  \vspace{.6cm}
\caption{The dashed lines in (a) represent the estimated posterior mean survival curves for the proposed BNP method. The dotdash lines in (b) represent the estimated posterior mean survival curves for the Naive method. In both figures, the solid lines represent the Kaplan-Meier curves of the observed survival data in control and treatment groups, and the dotted lines represent 95\% point-wise credible intervals of the posterior estimated survival curves. Survival times are on the original scale (days).}
\label{fig:surv}
\end{figure}

An initial analysis of the brain tumor death outcome using Kaplan-Meier is given in Figure \ref{fig:surv}, indicating that the treatment group has higher estimated survival probabilities.  The estimated difference at 365 days is 2.6\% (95\% CI: -8.1\% to 13.3\%).   
Figure \ref{fig:surv} plots the estimated posterior survival curves for treatment and control groups  marginalized over the distribution of covariate with 95\% credible intervals;  panels (a) and (b) display the results for the BNP and Naive approaches, respectively.  Using the BNP approach, the estimated posterior difference in survival at 365 days is 6.2\% (95\% CI: -1.2\% to 13.3\%).  For the Naive approach, the estimated posterior difference in survival at 365 days is 8.4\% (95\% CI: 0.2\% to 17.9\%).   The BNP approach produces comparable or higher treatment-specific estimates of survival and greater treatment differences than Kaplan-Meier.  In contrast, the Naive approach produces comparable or lower (higher) estimate of survival for the control (treatment) group than Kaplan-Meier.  Comparatively speaking, the Naive approach produces lower treatment-specific posterior estimates of survival than the BNP approach.  When we compare the fit to the observed survival data of the BNP and Naive approaches using the log-pseudo marginal likelihood (LPML) \citep{Geis:Eddy:1979}, a leave-one-out cross-validation statistic, we see the BNP performs better.  Specifically, the LPML for the treatment arm is -144 and -161 for the BNP and Naive approaches, respectively.  The corresponding numbers for the control arm are -137 and -174.

\begin{figure}
\begin{tabular}{cc}
  \includegraphics[scale=0.4]{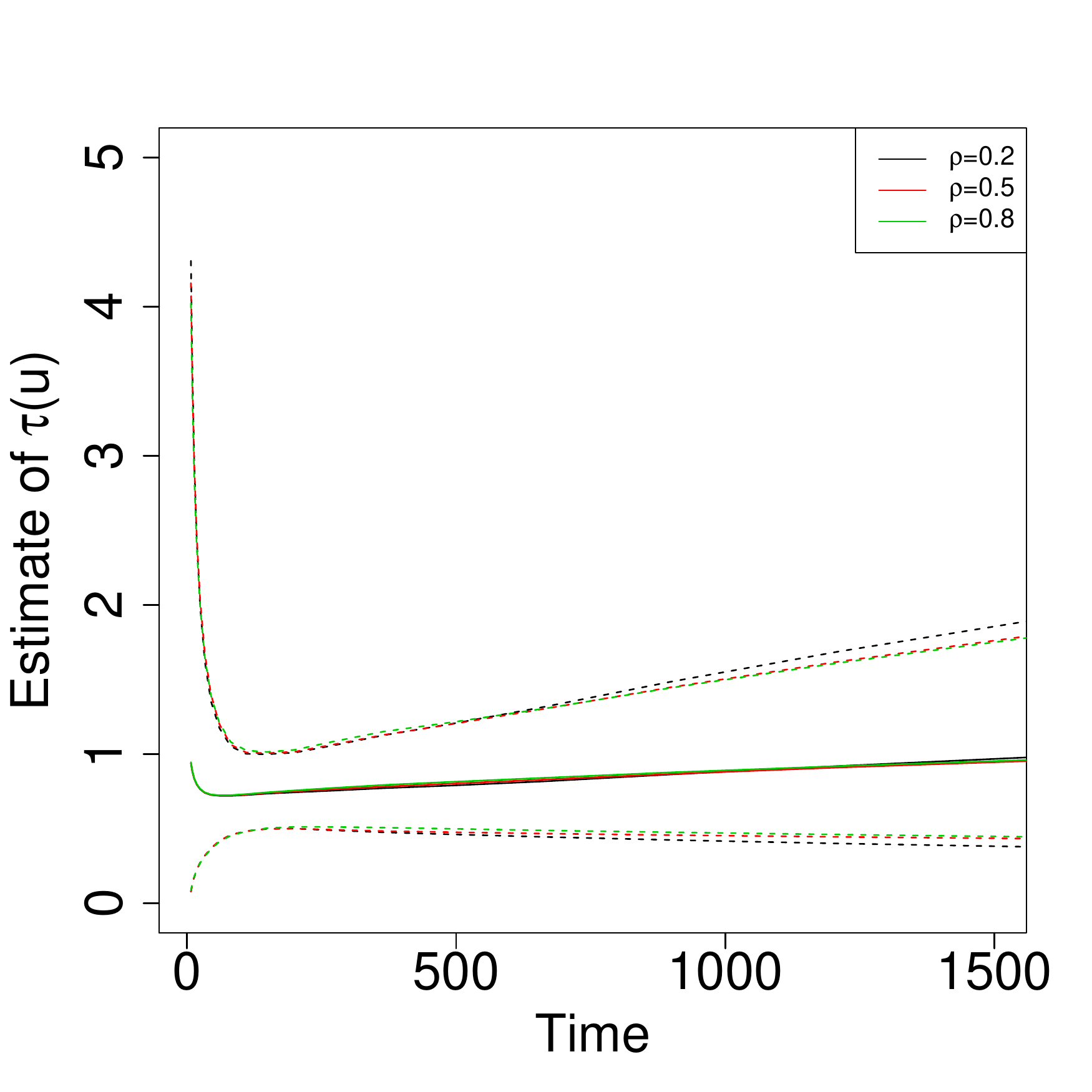}
  & \includegraphics[scale=0.4]{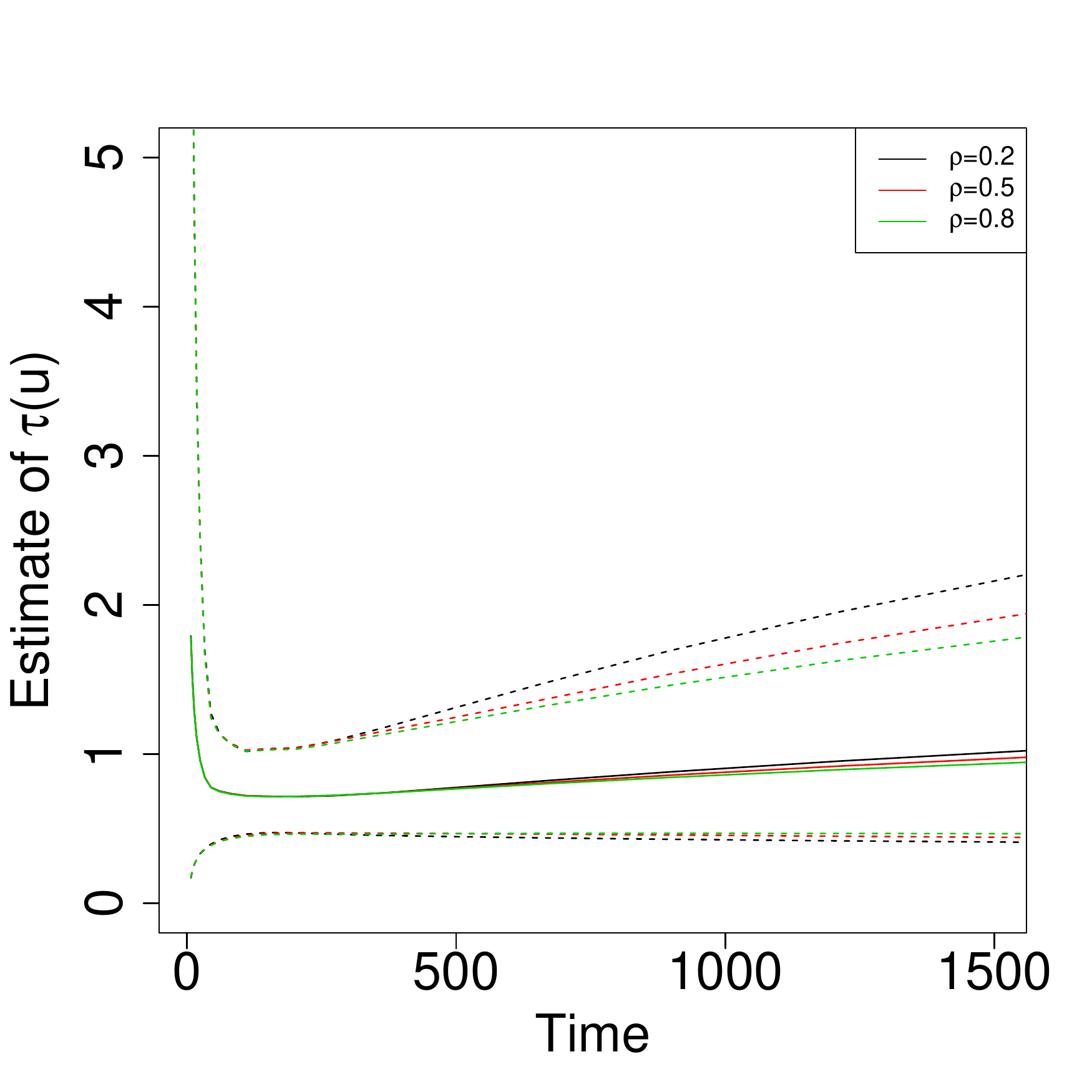}\\ 
(a) BNP&(b) Naive \\
\end{tabular}
  \vspace{.6cm}
\caption{Posterior estimated $\tau(u)$ versus $u$ on the original scale (days) in brain tumor data analysis for different $\rho$'s under the proposed BNP method and the Naive method, respectively. The solid lines represent the posterior estimated  $\tau(u)$, and the dashed lines represent 95\% point-wise credible intervals. }
\label{fig:brainh}
\end{figure}

For the BNP (see panel (a)) and Naive (see panel (b)) approaches, Figure  \ref{fig:brainh} plots the posterior estimates (along with point-wise 95\% credible intervals) of the causal estimand $\tau(u)$ versus $u$ for three choices of $\rho$, 0.2, 0.5 and 0.8.  Except near $u=0$, there are no appreciable differences between the two approaches.  In addition, the results are insensitive to choice of $\rho$.   Overall, this analysis shows that there is a lower estimated risk of progression for treatment versus of control at all time points, except near zero.  However, there is appreciable uncertainty, characterized by wide posterior credible intervals, that precludes more definitive conclusions about the difference between treatment groups with regards to progression.   When we compare the fit to the observed survival {\em and} progression data of the BNP and Naive approaches using LPML,
we see that the approaches perform comparably.  Specifically, the LPML for the treatment arm is -227 and -232 for the BNP and Naive approaches, respectively.  The corresponding numbers for the control arm are -215 and -214.

\section{Discussion}
\label{sec:con}

In this paper, we proposed a  causal estimand for characterizing the effect of treatment on progression in a randomized trials with a semi-competing risks data structure.  We introduced a set of identification assumptions, indexed by a non-identifiable sensitivity parameter that quantifies the correlation between survival under treatment and survival under control.   Selecting a range of the sensitivity parameter $\rho$ in a specific trial will depend on clinical considerations. For example, in trial of a biomarker targeted therapy, one might expect weaker correlation, since survival under control might be primarily determined by co-morbidities and the survival under treatment might be more determined by the presence of the targeted molecular aberration. In contrast, for some chemotherapies, the same factors that impact survival under control may equally impact survival under treatment, e.g., co-morbidities, social support, etc.  Fortunately, the sensitivity parameter is bounded between -1 and 1 and, in most settings, should be positive; a range should be selected in close collaboration with subject matter experts.

We proposed a flexible Bayesian nonparametric approach for modeling the distribution of the observed data.  Since the causal estimand is a functional of the distribution of the observed data and $\rho$, we draw inference about it using posterior summarization.  Our procedure can easily be extended to accommodate a prior distribution on $\rho$, which 
will allow for integrated inference.  Our procedure also allows for posterior inferences about other identified causal contrasts such as the distribution of survival under treatment versus under control.  The procedure can also be used for predictive inference for patients with specific covariate profiles.

\section*{Acknowledgements}
This research is supported by NIH CA183854 and GM 112327. The authors would like to thank Drs. Henry Brem and Steven Piantadosi for providing access to data from the brain cancer trial.

\bibliographystyle{biorefs}
\bibliography{semi,BNP}
\clearpage
\newpage

\section*{Appendix}

 \subsection*{A1: Determining Prior Hyperparameters}
As priors for $\bbeta^z_{hj}$ in
 the GP mean function, we assume 
 $\bbeta^z_{hj} \sim N(\bbeta^z_{0j}, \bLambda^z_{0j})$. We assume $\bLambda^z_{0j} \sim \IW(\lambda^z_0, \bPsi^z)$, where
$E[\bLambda_{0j}^z]=\frac{ \bPsi^z}{\lambda^z_0-3}$. 
The precision parameter $\alpha$ in the DDP is assumed to be distributed $\mathrm{Ga}(\lambda_1, \lambda_2)$.
 
In  applications of Bayesian inference with small to moderate sample sizes, 
a critical step is to fix values for all hyperparameters $\omega = \{
\bbeta^z_{0j},   \bSig^z_{0j}, \lambda_0^z, \Phi^z, z=0, 1, j=0, 1,
\lambda_1, \lambda_2 \}$.  Inappropriate information could be
introduced by improper numerical values, leading to inaccurate posterior
inference.  We use an empirical Bayes method to obtain
$\bbeta^z_0=(\bbeta^z_{01}, \bbeta^z_{02})$ by fitting a bivariate 
normal distribution for responses of patients under treatment $z$,
$\bY\mid Z=z \sim N(\bx\bbeta^z_{0},
\bSigma_{\bbeta_0})$. For $\bSig^z_{0j}$, we assume a diagonal matrix
with the diagonal values being 10. After an empirical estimate of
$\hat{\Sigma^z}$ is computed, we tune $\lambda_0^z$ and  $\Phi^z$ so
that the prior mean of $ \bSigma^z$ matches the empirical estimate,
$\lambda_0^z=4$ and   
 $\Phi^z=\hat{\bSigma^z}$. Finally, we assume $\lambda_1=\lambda_2=1$. 

\subsection*{A2: MCMC Computational Details}

Unless required for clarity, we suppress dependence of the notation on treatment $z$.  Here $j$ is used to denote endpoint ($j=1$ for progression and $j=2$ for death).  We define
\[
\bSigma = \left[  \begin{array}{ll} \sigma^2_1 & \sigma_{12} \\ \sigma_{21} & \sigma^2_{2} \end{array}  \right].
\]
Let $A_h = \{ i : \gamma_i = h \}$ and $n_h = | A_h |$, $\bY_i = (Y_{Pi}, Y_{Di})$, $\bths_{hj}=(\th_{hj}(\bX_1),\ldots,\th_{hj}(\bX_n))$ ($h = 1,\ldots, K$), $\bX$ is an $n \times D$ matrix where the $i$th row contains the $D$-dimensional covariate vector $\bX_i$ for the $i$-th patient, $\bR_j$ is an $n \times n$ matrix where the $(i,i')$ entry is $R_j(\bX_i,\bX_{i'})$, $\bU_h$ is an $n_h \times n$ matrix where the $k$th row refers to the $k$th patient in $A_h$, the $i$th column refers to patient $i$ and $(k,i)$ element is the indicator that the patient in $k$th row is the same as the patient in the $i$th column, $\bI_n$ is an $n \times n$ identity matrix, $\tilde{\bY}_{hj} = \{ \tilde{Y}_{ji} : \gamma_i=h \}$ where $\tilde{Y}_{1i} =Y_{Pi} - \frac{\sigma_{12}}{\sigma_2^2} \left( Y_{Di} - \theta_{\gamma_i 2}(\bX_i) \right)$ and $\tilde{Y}_{2i} =Y_{Di} - \frac{\sigma_{21}}{\sigma_1^2} \left( Y_{Pi} - \theta_{\gamma_i 1}(\bX_i) \right)$, $\tilde{\sigma}_1^2 = \sigma_1^2 - \frac{\sigma_{12} \sigma_{21}}{\sigma_2^2}$ and $\tilde{\sigma}_2^2 = \sigma_2^2 - \frac{\sigma_{12} \sigma_{21}}{\sigma_1^2}$.

For $z=0,1$, we iterate through the following 6 updating steps:

\begin{enumerate}

%

\item Update $w_h$
\begin{eqnarray*}
v_h\sim Beta(1+n_h, \alpha+\sum_{j>h}n_j),
\end{eqnarray*}
where $n_h=\sum_{i=1}^{n_k}I(\gamma_i=h)$ is the number of observations such that $\gamma_i=h$.
Then $w_1=v_1$ and $w_h=v_h\prod_{j<h}(1-v_j)$.

\item{Update $\alpha$}

Assuming that $\alpha\sim \mathrm{Ga}(\lambda_1, \lambda_2)$, 
$$\alpha \sim \mathrm{Ga}(\lambda_1+H-1, \lambda_2-\sum_{h=1}^{H-1}\log(1-v_h)),$$
where $v_h$ is generated from step 1.

\item{Update $\bSigma$}
\[
\bSigma\mid \cdot \sim \mathrm{Inverse \ Wishart} \left( \lambda_0+n, \bPsi+\sum_{h=1}^K\sum_{i: \gamma_i=h}(\bY_i-\bth_h(\bX_i))(\bY_i-\bth_h(\bX_i))' \right)
\]

\item Update $\bths_{hj}$, $j=1, 2$
\begin{eqnarray*}
p(\bths_{hj}\mid \cdot)&\propto& p(\bths_{hj})\prod_{i:\gamma_i=h}p(\bY_i\mid \btheta_{h}(\bX_i)) \nonumber\\
&\propto& \exp\left\{-\frac{1}{2}(\bths_{hj}-\bX\bbeta_{hj})' \bR_j^{-1}(\bths_{hj}-\bX\bbeta_{hj})\right\} \nonumber\\
&&\times \exp \left\{-\frac{1}{2}\sum_{i:\gamma_i=h}(\bY_i-\btheta_h(\bX_i))' \bSigma^{-1} (\bY_i-\btheta_h(\bX_i)) \right\} \nonumber\\
&\sim&\mathrm{N} \left( \left\{ \bR_j^{-1}+\frac{\bU_h'\bU_h}{\tilde{\sigma}_j^2} \bI_{n} \right\} ^{-1} \left\{ \bU_h'\frac{ \tilde{\bY}_{hj}}{\tilde{\sigma}_j^2}+\bR_j^{-1}\bX\bbeta_{hj} \right\}, \left\{ \bR_j^{-1}+\frac{\bU_h' \bU_h}{\tilde{\sigma}_j^2} \bI_{n} \right\}^{-1} \right)\nonumber\\\end{eqnarray*}
\item Update $\bbeta_{hj}$
\begin{eqnarray*}
p(\bbeta_{hj}\mid \cdot)&\propto& p(\bbeta_{hj})  \exp\left\{-\frac{1}{2}(\bths_{hj}-\bX\bbeta_{hj})' \bR_j^{-1}(\bths_{hj}-\bX\bbeta_{hj})\right\} \nonumber\\
&\sim& \mathrm{N}(\bLambda_{hj} \left( \bX' \bR_j^{-1}\bths_{hj}+\bLambda^{-1}_{0j}\bbeta_{0j} \right), \bLambda_{hj}),
\end{eqnarray*}
where $\bLambda_{hj} = (\bX' \bR_j^{-1}\bX + \bLambda_{0j}^{-1})^{-1}$.

\item Update $(\gamma_i, \bY_i)$, where $\bY_i = (Y_{Pi},Y_{Di})$.  

We write $p(\gamma_i, \bY_i \mid \cdot)$ as $p(\bY_i \mid \gamma_i, \cdot) p(\gamma_i \mid \cdot)$ where $\cdot$ includes $\bO_i$.  

\begin{itemize}
\item If $\delta_i=\xi_i=1$ 
\begin{eqnarray*}
Pr(\gamma_i=h\mid \cdot)&\propto& w_h \phi(T_{1,i},T_{2,i}; \btheta_h(\bX_i), \bSigma); 
\end{eqnarray*}
$p(\bY_i \mid \gamma_i, \cdot)$ is point mass at $\bY_i = (T_{1i},T_{2i})$.
\item If $\delta_i=\xi_i=0$ (i.e., $Y_{Pi} > T_{2i}, Y_{Di} > T_{2i}$), 
\begin{eqnarray*}
Pr(\gamma_i=h\mid \cdot)&\propto& \int_{t > T_{2i}}  \int_{s > T_{2i}} w_h \phi(s,t; \btheta_h(\bX_i), \bSigma)ds dt.
\end{eqnarray*}
\[
p(\bY_i \mid \gamma_i, \cdot) = \frac{\phi( \bY_i; \btheta_{\gamma_i}(\bX_i), \bSigma)}{ \int_{t > T_{2i}}  \int_{s > T_{2i}} \phi(s,t; \btheta_{\gamma_i}(\bX_i), \bSigma)ds dt },
\]
where $Y_{Pi} > T_{2i}, Y_{Di} > T_{2i}$.
\item If $\delta_i=1$ and $\xi_i=0$ 
\begin{eqnarray*}
Pr(\gamma_i=h\mid \cdot)&\propto& \int_{t > T_{2i}} w_h\phi(T_{1i},t; \btheta_h(\bX_i), \bSigma) dt;
\end{eqnarray*}
\[
p(\bY_i \mid \gamma_i, \cdot) = \frac{\phi( \bY_i; \btheta_{\gamma_i}(\bX_i), \bSigma)}{  \int_{t > T_{2i}} \phi(T_{1i},t; \btheta_{\gamma_i}(\bX_i), \bSigma) dt } ,
\]
where $Y_{Pi} = T_{1i}, Y_{Di} > T_{2i}$.
\item If $\delta_i=0$ and $\xi_i=1$ 
\begin{eqnarray*}
Pr(\gamma_i=h\mid \cdot)&\propto& \int_{s > T_{1i}} w_h\phi(s,T_{2i}; \btheta_h(\bX_i), \bSigma) ds;
\end{eqnarray*}
\[
p(\bY_i \mid \gamma_i, \cdot) = \frac{\phi( \bY_i; \btheta_{\gamma_i}(\bX_i), \bSigma)}{ \int_{s > T_{1i}} \phi(s,T_{2i}; \btheta_{\gamma_i}(\bX_i), \bSigma) ds }, 
\]
where $Y_{Pi} > T_{1i}, Y_{Di} = T_{2i}$.
\end{itemize}


\end{enumerate}

\newpage

\newpage

\newpage

\clearpage
\newpage

\begin{table}
\centering
  \begin{tabular}{|c|c|c|c|l|}
    \hline
    \multirow{2}{*}{Scenario} &
      \multicolumn{2}{c|}{$Z=0$} & \multicolumn{2}{c|}{$Z=1$}  \\ \cline{2-5}
    & BNP & Naive & BNP &Naive\\
    \hline
    1 & 0.012 (0.007) &0.013 (0.007)  & 0.012 (0.006) & 0.013 (0.007)   \\
    \hline
    2 & 0.042 (0.022) & 0.088 (0.032)  & 0.019 (0.007) &0.073 (0.035)  \\
    \hline
    3 & 0.012 (0.006) & 0.013 (0.007)  &0.012 (0.007)&  0.014 (0.007)  \\
    \hline
  \end{tabular}
  \vspace{.6cm}
  \caption{For each scenario, mean and standard deviation of RMSE across 500 simulated datasets under the proposed BNP method and the naive Bayesian method (Naive). }
  \label{table:simu}
\end{table}

\clearpage
\newpage

\begin{table}
\centering
{\footnotesize
  \begin{tabular}{|c|c|c|c|c|c|c|c|}
    \hline
    \multirow{2}{*}{Scenario} &
      \multicolumn{2}{c|}{$\rho=0.2$} &
      \multicolumn{2}{c|}{$\rho=0.5$} &
       \multicolumn{2}{c|}{$\rho=0.8$} 
       \\ \cline{2-7}
    & BNP & Naive &BNP & Naive&BNP & Naive \\ 
    \hline
    1 &0.286 (0.087)& 0.328 (0.126)& {\bf 0.059 (0.035)} & 0.073 (0.051)& 0.185 (0.037)& 0.207 (0.047)   \\ 
    \hline
    2 & 0.277 (0.128) & 0.493 (0.250) & {\bf 0.090 (0.062)}& 0.199 (0.169) & 0.261 (0.070) & 0.243 (0.111) \\
    \hline
    3 & 0.106 (0.032) & 0.105 (0.038) &{\bf 0.033 (0.016)}& 0.035 (0.021)& 0.086 (0.028) & 0.097 (0.034)  \\
    \hline
  \end{tabular}  }
    \vspace{.6cm}
  \caption{Means and standard deviations of RMSE for estimating $\hat{\tau}(u)$ across 500 simulations in three scenarios under the proposed BNP approach and naive Bayesian method (Naive), respectively. }
  \label{table:simuh}
\end{table}

\label{lastpage}

\end{document}